\documentclass[runningheads]{llncs}
\usepackage[T1]{fontenc}
\usepackage{graphicx}
\newcommand{\N}{\mathrm{I\kern-.23emI\kern-.29em N}}
\newcommand{\R}{\mathrm{I\kern-.23emI\kern-.29em R}}
\begin{document}
\title{Online Competitive Searching for Rays\\ in the Half-plane}
%
%
\author{ Elmar Langetepe \and Florian Gans  }
\authorrunning{E. Langetepe and F. Gans }
%
\institute{ University of Bonn, Institute of Computer Science V, D-53113 Bonn,
            Germany \email{\{elmar.langetepe,s6flgans\}@uni-bonn.de}}

\maketitle              

\begin{abstract}
We consider the problem of searching for rays (or lines) in the half-plane. The given problem turns out to be a very natural extension of the cow-path problem that is lifted into the half-plane and the problem can also directly be motivated by a  $1.5$-dimensional terrain search problem. We present and analyse an efficient strategy for our setting and guarantee a competitive ratio of less than $9.12725$ in the worst case and also prove a lower bound of at least $9.06357$ for any strategy. Thus the given strategy is almost optimal, the gap is less than $0.06368$. By appropriate adjustments for the terrain search problem we can improve on former results and present geometrically motivated proof arguments. As expected, the terrain itself can only be helpful for the searcher that competes against the unknown shortest path. We somehow extract the core of the problem.

\end{abstract}
%
%
%

\section{Introduction}\label{sect-intro}
\begin{figure}
\includegraphics[width=\textwidth]{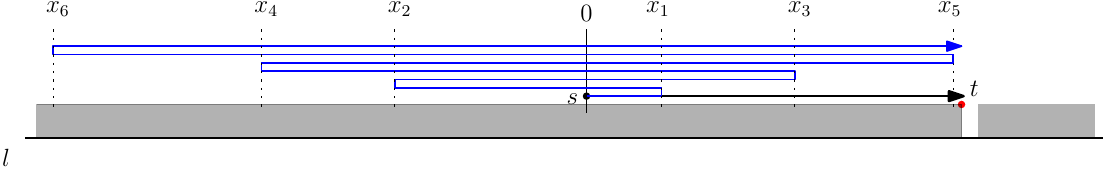}
\caption{The \emph{cow} is searching for the unknown \emph{hole in the fence} and  moves forth and back along the fence.  Any reasonable search strategy can be represented by a sequence $X=(x_1,x_2,x_3,x_4,\ldots)$.  The cow  runs into a local worst case situation in relative comparison to the unknown shortest path to the target, when the target is distance $|st|=x_k+\epsilon$ away from the start and a full turn on the other side up to distance $x_{k+1}$ was done before (here for $k=5$).}\label{fig-cowpath}
\end{figure}

In the online search game and online algorithms community the so called cow-path problem is a well-known and very famous search problem.  The \emph{cow} is located  close to a fence and is searching for a hole that is a certain distance away. The cow has no further idea where the hole might be, how far it is away and it is also dark night, see Figure~\ref{fig-cowpath}.
 Geometrically speaking a searcher is located on a fixed starting position~$s$ on an (infinite) line~$l$ and is searching for an unknown (immobile) target point~$t$ somewhere on~$l$. If the agent knows the position of the target $t$, she can move directly toward it with optimal minimal search cost $|st|$ on the line.  Without further information or vision any reasonable search path has to visit both directions starting from $s$ in an alternating fashion running forth and back  with successively increasing search depth for both possible directions. Thus any reasonable strategy  can be described by an (infinite) sequence of moves $X=(x_1,x_2,x_3,x_4,x_5,\ldots)$, $x_i\in \R$ for $i\in \N$, such that the agent visits one direction up to increasing depth $x_1<x_3<x_5<\ldots$ and in the other direction up to increasing depth $x_2<x_4<x_6<\ldots$ and visits the two directions alternately. Note that the agent moves back to the start, so any distance is counted twice until the goal is met. We would like to compare the path length of a successful search strategy  to the shortest path to the target under full information. For a given strategy~$X$ a (local) worst case situations in relative comparison to the unknown shortest path to the target  appears, when the agent slightly miss the target at step $k$ at distance $x_k$, starts a new iteration to the other side with distance $x_{k+1}$ and after turning back, the target is detected closely behind the former visit at some distance $|st|=x_k+\epsilon$ for an arbitrarily small~$\epsilon$. We are searching for an efficient \emph{online}  strategy under incomplete information against an optimal \emph{offline} path under full information, in this case the direct shortest path to the goal. 

This is the general notion of competitive analysis for \emph{online} algorithms or strategies. 
W.r.t. to an efficient online approximations one is looking for a sequence~$X$ that guarantees that the length of the online search path to a goal is only~$C$ times worse than the shortest path to this goal. Therefore in this setting we would like to find a sequence such that for a constant~$C\geq  1$ we guarantee 
\begin{equation}\label{equ-worst-case}
2\sum_{i=1}^{k+1} x_i+(x_{k}+\epsilon)\leq C\cdot (x_{k}+\epsilon) \mbox{ holds for any } k\in \N \mbox{ and any } \epsilon>0\,.
\end{equation}
We would like to find the optimal strategy $X$ that minimizes $C$ under all such constants.  
Note that the factor~2 of the sum in (\ref{equ-worst-case}) comes from moving forth-and-back. For $k=0$ there is no $x_0$ (or set $x_0:=0$ as an alternative). This  means that such a constant $C$ cannot exists, since for any fixed $x_1>0$ and $C$ we can make use of  an arbitrarily small $\epsilon$ such that $2x_1 +\epsilon\leq C\cdot \epsilon$ does not hold. We overcome such a starting situation by assuming that the goal is at least one step away from the start (say $x_0:=1$). Alternatively an additive constant in the notion of \emph{competitive analysis}, see Section~\ref{sect-model} can be used. 
 In any case we can also neglect $\epsilon$ and we are searching for a sequence that minimizes the following ratio and guarantees a constant $C$ such that
$$ 
\frac{2\sum_{i=1}^{k+1} x_i+x_{k}}{x_k}= 1 +2\frac{\sum_{i=1}^{k+1} x_i}{x_k}  \leq C \mbox{ holds for any } k\in \N\,. 
$$
Note that it is obvious that the above local worst case guarantee is sufficient to guarantee the approximation ratio~$C$ for all other targets. For example while moving from $x_k$ to $x_{k+2}$ any potential target will result in a better ratio. 

The problem was first introduced as a linear search problem by Beck~\cite{beck1964linear} in 1964. Since then it was reconsidered in many aspects.  
It is well-known that a doubling heuristic with $x_i=2^{i-1}$ give  an optimal strategy with $\frac{\sum_{i=1}^{k+1} 2^{i-1}}{2^k} =4-\frac{1}{2^k}\leq 4$ and  $C=9$. Different proofs were given for the optimality result for this constant. The first proof was given by Gal~\cite{g-sg-80} in the context of search games and by the famous generic result on optimizing functionals, i.e.,  $F_k(X):=\frac{\sum_{i=1}^{k+1} x_i}{x_k}$ is an example of a functional defined over a sequence~$X$. Later the problem was re-invented by Baeza-Yates et. al~\cite{bcr-sp-93}. 

\subsection{Ray search in the half-plane}\label{subsect-intro}
\begin{figure}
\includegraphics[width=\textwidth]{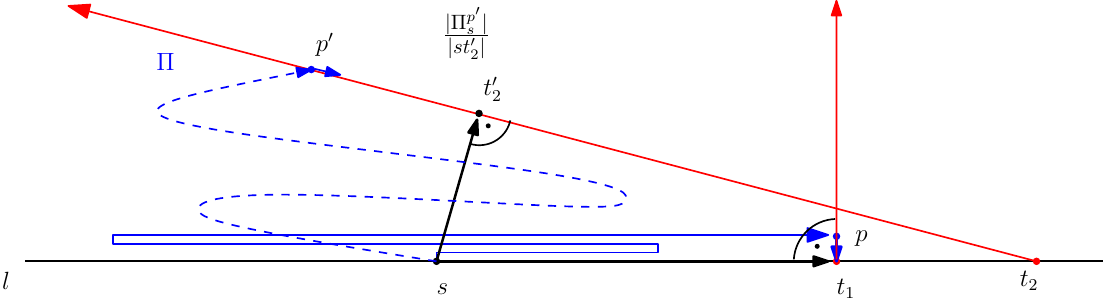}
\caption{Searching for an unknown  ray in the half-plane and competing with a search path $\Pi$ against the shortest path to the ray. If the goal set is restricted to rays perpendicular to $l$, we will simply get back to the cow-path problem.}\label{fig-rayscenario}
\end{figure}

In this paper we lift the cow path problem into the half-plane and consider the following task. Starting from some point $s$ on a line $l$ in the Euclidean 2D plane we are searching for a ray $r$ that has its starting point $t$ on $l$  and points into one half-plane w.r.t. $l$. W.l.o.g. we assume that~$l$ is the~$X$-axis of the coordinate system,~$s$ is the origin $(0,0)$ and the rays $r_t$ point into the upper half-plane w.r.t. the $X$-axis with a start point $t=(t_x,0)$. An unknown ray will be detected by a search path when the ray is visited for the first time, see Figure~\ref{fig-rayscenario}. Only searching for rays that run perpendicular to the $X$-axis directly results in the original cow-path problem, because there is no benefit in moving into the half-plane. 
For more general rays a search path $\Pi$ has to move into the half-plane. If an unknown ray will be visited by $\Pi$ for the first time at some point $p'$, we consider the ratio of the length of the search path, $|\Pi_s^{p'}|$ from $s$ to $p$ over the shortest path to the ray which is a segment $s t'$ that runs perpendicular to the ray. 

\begin{figure}[h]
\begin{center}
\includegraphics[scale=0.5]{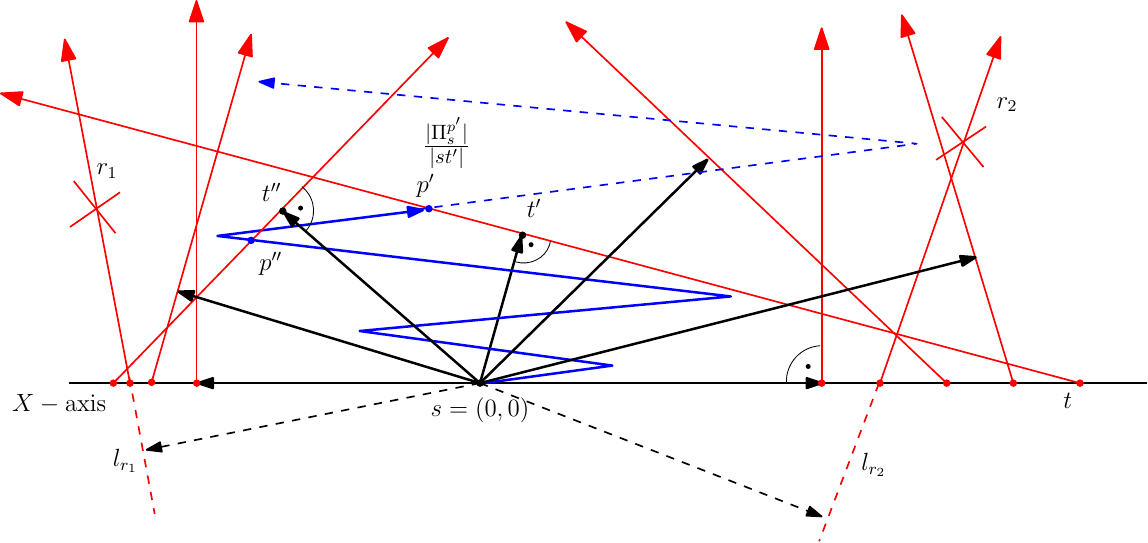}
\caption{An example subset of rays we are looking for. For any such ray $r$ the shortest path from $s$ to the ray, i.e., a segment $st'$ with $t'\in r$ that runs perpendicular to $r$, also has to lie in the upper half-plane. 
Here the rays $r_1$ and $r_2$ are not in our focus. A reasonable search path, $\Pi$, might be given by a generalized cow-path that also increases its height successively. Note that we compete against any possible target ray with the given property.}\label{fig-rayall}
\end{center}
\end{figure}

We would like to make use of search paths that are generalized cow-path strategies acting in the upper half-plane with increasing search depth to the left and to the right and also with increasing depth in the $Y$-direction. Therefore we restrict the rays that we are looking for as follows. This restriction is also well motivated by the terrain search application mentioned in the next section. Any ray $r$ with source point $t$ on the $X$-axis can be extended to a line $l_r$. The search path will run in the upper half-plane. To be close to the cow-path problem we require that the shortest path to any indicated $l_r$, i.e., a line segment $st'$ that runs perpendicular to $l_r$ also has to be keep inside this half-plane.  
In other words the corresponding rays \emph{point} toward the origin~$s$. Rays with a source to the left of $s$ have positive slope up to $\infty$ and rays with a source to the right of $s$ has negative slope up to $-\infty$. Now any reasonable (infinite) search path, $\Pi$, has to visit all such rays, moves from left to right and with increasing search depth in both $X$-directions and alos in the $Y$-direction, see Figure~\ref{fig-rayall}. 

In this paper we first present a (cow-path related) strategy in 2D that guarantees a competitive ratio of less than $9.12725$ against the shortest path to any unknown ray in the half-plane that points toward the origin. We also present a lower bound of~$9.06225$. So the gap has size strictly less than $0.065$. The restriction to rays (or lines) pointing toward the start is well motivated by the terrain search problem discussed in Section~\ref{subsect-terrain}. We also adjust our strategy to the terrain search problem by an appropriate modification. Fortunately, by geometric arguments we can show that this modification is sufficient to guarantee the presented ratios for the terrain also. Obviously (as expected) the terrain just gives us an advantage. Altogether we improve on the above former results. Our different approach for solving this problem also somehow extracts the core of the terrain search problem. This should make it easier to potentially find its overall competitive complexity. 

\section{Related work and terrain search application}\label{sect-model}

Analytically we will make use of the competitive framework which goes back to Sleater and Tarjan \cite{sleator1985amortized} and was first used for list update and paging problems. 
Let us consider a specific algorithmic problem and let $\mathcal{P}$ be the set of all instances of the problem. An optimal offline algorithm \emph{OPT} solves the problem and runs under full information. Furthermore, the cost of such an algorithm for an instance  $P\in \mathcal{P}$ is denoted as $S_{OPT}(P)$ and optimality means that \emph{OPT} attains the minimal cost for any instance and among any algorithm. Let us further assume that an online algorithm \emph{ALG} also solves any problem instance of  $P\in \mathcal{P}$ but attains the problem information successively. Thus \emph{ALG} might not solve the problem with minimal cost, i.e.,  $S_{ALG}(P)\geq S_{OPT}(P)$.  We call the online algorithm $C$-competitive if there are constants $A$ and $C$ such that the inequality 
\begin{equation}\label{equ-compAn}
S_{ALG}(P)\leq C \cdot S_{OPT}(P) + A
\end{equation}
holds for all $P\in  \mathcal{P} $. So \emph{ALG} guarantees to be no worse than~$C$ times the optimal offline solution, apart from an additive constant~$A$. This second constant~$A$ covers starting situations. It is required when for some fixed small instances some first steps of the online algorithm cause relatively high cost against the optimum. In this case the problem might not be analysed properly. Thus it is allowed that small instances or special starting situations can be fully covered or excluded by additive constants. It is therefore sufficient to consider the ratio $\frac{S_{ALG}(P)}{S_{OPT}(P)}$ for analysing a specific strategy. For lower bounds the constant $A$ has to be taken into account. Thus a lower bound construction requires the option of guaranteeing arbitrarily large values for $S_{OPT}(P)$ (and thus in turn also for $S_{ALG}(P)$). 

The mentioned cow-path problem has a long tradition. And  many further extensions and modifications already have been discussed. The problem was first mentioned as a linear search problem by Beck~\cite{beck1964linear} in 1964 and he came back to the problem later~\cite{beck1970yet}. In the context of search games optimal solutions have been first given by Gal~\cite{g-sg-80}. The problem was introduced to the computational geometry community by Baeza-Yates et. al~\cite{bcr-sp-93}. The effect of randomization was considered~\cite{kao1996searching},  the problem was extended to $m$-rays~\cite{lopez2001ultimate} and also considered w.r.t. turn costs~\cite{demaine2006online} and bounded distances~\cite{hipke1999find}. Some generic characterization of optimal strategies have been found~\cite{bose2015searching,lopez2001ultimate}. Recently also the concept of hints for distances and directions have been discussed~\cite{angelopoulos2023online}. Extending the search problem to searching for lines (or rays) in 2D is an ultimate challenge. The main conjecture is that a logarithmic spiral is the overall best competitive strategy (for searching for lines in the plane) with a proven competitive ratio of~$13.8111351795\ldots$ This is also known as the \emph{swimming-in-the-fog} or \emph{lost-in-a-forest} problem, i.e., searching for a (shore)line in the plane; see the overview by Chrobak~\cite{chrobak2004princess} for the case when the distance to the line is known, the famous \emph{textbook} of Alpern and Gal~\cite{alpern2003theory} on search games and rendezvous and the report of Finch~\cite{finch2005logarithmic}. Up to know the optimality of spiral search in the plane was only shown in the context of radar-search for points in the plane~\cite{langetepe2010optimality}. Some attempts were made for the restriction on axis-parallel shorelines~\cite{jez2009two,langetepe2012searching}.

\subsection{Terrain search application}\label{subsect-terrain}

The most specific motivation of the problem discussed here stems from a recent work from S.~de~Berg et al. (Latin 2024)~\cite{de2024competitive}. The authors consider an unknown 1.5D terrain $T$, the upper boundary is modelled by a $X$-monotone (polygonal chain) function in the Euclidean coordinate system. More precisely the upper boundary of the terrain has to be $X$-monotone or monotone w.r.t. some ground level $G$ that can be assumed to run in parallel to the $X$-axis. Guarding of 1.5D terrains also has attracted some attention before, for example the terrain was formerly considered w.r.t. computing the number of guards required~\cite{ben2007constant,elbassioni2011improved}. 

\begin{figure}[h]
\begin{center}
\includegraphics[scale=0.35]{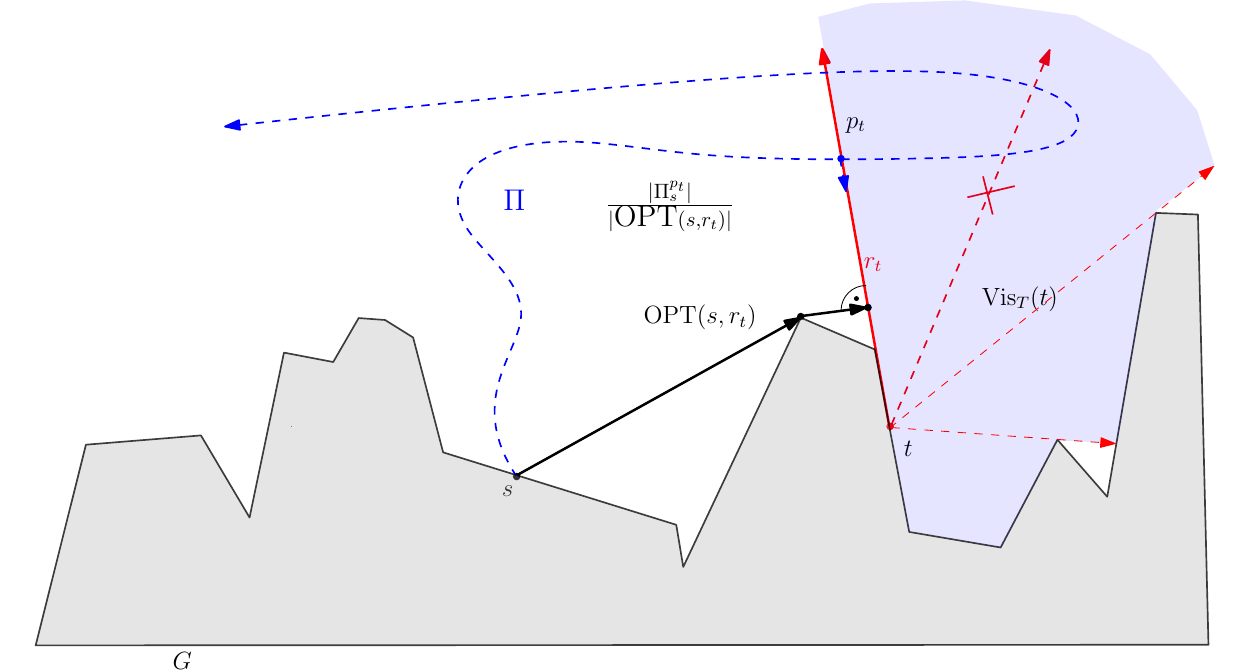}
\caption{Searching for an unknown target $t$ in a 1.5D terrain $T$. By entering the visibility region, $\mbox{Vis}_T(t)$, of $t$ for the first time a ray $r_t$ that points toward the start will be visited first. Visibility rays pointing away from the start need not be considered. The shortest path for visiting the ray $r_t$ from $s$ that avoids the terrain, $\mbox{OPT}_T(s,r_t)$, need not be a single line segment.}\end{center}\label{fig-terrainexpl}
\end{figure}

Here an agent starts somewhere on the terrain and has to keep above the terrain along a search path, $\Pi$, in order to find an unknown target $t$ located on the terrain, comparable to  a helicopter that is moving above the terrain,  see Figure~\ref{fig-terrainexpl}. Along the search path $\Pi$ at some point $p_t$ the target might get into sight for the first time, i.e., the segment $t\,p_t$ does not cross the terrain any more. In this case the target $t$ will be detected.  Formally, at $p_t$ the searcher enters the visibility region of the target $t$, $\mbox{Vis}_T(t)$,  for the first time. By the $X$-monotonicity of the terrain boundary, this will indeed always occur for a ray $r_t$ that \emph{points} toward the target. Now the searcher competes against the shortest path from the start $s$ that has to keep above the terrain and also visits this ray, i.e., the offline shortest path for detecting the goal, $\mbox{OPT}_T(s,r_t)$.

Note that the restriction to special boundary functions ($X$-monotone polygonal chains or functions) is a necessary condition for reasonable approximations. If it would be allowed to use an arbitrary polygonal chain for the upper boundary of the terrain, one can easily model (simulate) arbitrary polygons along the boundary. This would mean that it will no longer be possible to achieve a constant competitive online approximation for arbitrary unknown targets in this setting. For the same reason the online agent needs some orientation, the agent need to have knowledge about the fixed ground level line~$G$. 

\begin{figure}[h]
   \begin{minipage}[]{.44\linewidth} 
      \includegraphics[width=\linewidth,page=1]{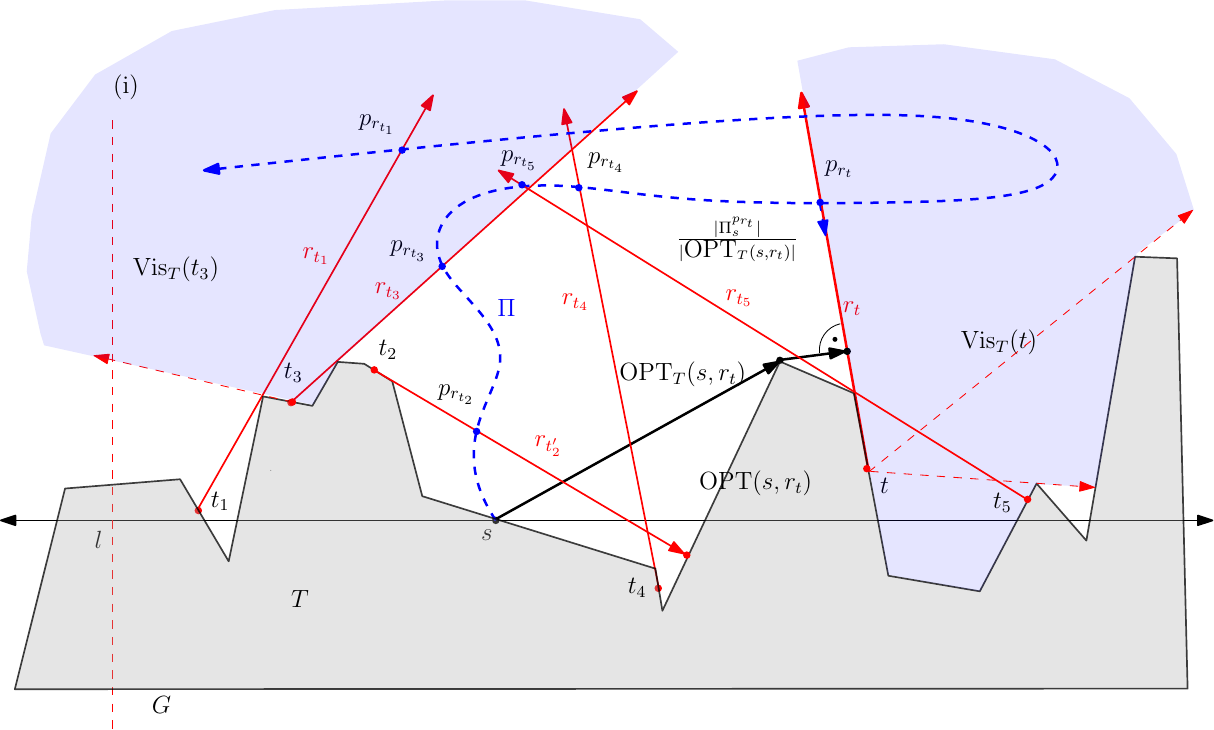}
   \end{minipage}
   \hspace{.1\linewidth}
   \begin{minipage}[]{.44\linewidth} 
      \includegraphics[width=\linewidth,page=2]{Terrain-Search-Scenario-Explain-Ext}
   \end{minipage}
    \begin{minipage}[]{.44\linewidth} 
      \includegraphics[width=\linewidth,page=3]{Terrain-Search-Scenario-Explain-Ext}
   \end{minipage}
   \hspace{.1\linewidth}
   \begin{minipage}[]{.44\linewidth} 
      \includegraphics[width=\linewidth,page=4]{Terrain-Search-Scenario-Explain-Ext}
   \end{minipage}
   \caption{(i) Detecting unknown targets $t$ on the terrain by a search path $\Pi$  means entering the visibility region $\mbox{Vis}_T(t)$ at a \emph{first} visibility ray $r_t$. All such rays \emph{point} toward the target $s$. (ii) We can now shift the target points to the line $l$ that runs through $s$ in parallel to the ground line $G$ of the terrain. (iii) Considering the corresponding rays for the new targets and omitting the overall terrain itself results in an instance of our (pure) ray-search problem. The shortest path to the target ray is a line segment and can only decrease. We compete against a better offline solution. (iv)  Any such ray-search scenario with a fixed set of rays can be easily transformed to a simple terrain search problem by making use of small pikes close to the target points.}\label{fig-terrainred}
\end{figure}

It is now easy to see that any such terrain configuration will induce a legal (subset) input for our, say pure, ray search problem in the half-plane. We explain the transformation which is not meant as a full reduction, it is meant as a motivation for our scenario, see Figure~\ref{fig-terrainred} for an overall example.
 To this end we first consider potential target points $t$ on the terrain that are not \emph{visible} from the start. Now for any $\mbox{Vis}_T(t)$ and any search path $\Pi$ there is a unique first ray, $r_t$, that will be visited by any given search path, $\Pi$, that starts from $s$ first, see Figure~\ref{fig-terrainred}(i).
This unique ray will also be visited by the optimal offline search path first at some point $p_{r_t}$. Thus actually only rays has to be detected. Additionally, all such potential rays obviously has to point toward the start in the sense mentioned before. There might be some exceptions at this point of our transformation, when the target lies on one side w.r.t. $s$ and the ray  points downwards to the other side of the start, see $t_2$ and $r_{t_2}$ in Figure~\ref{fig-terrainred}. This will cause no harm because we can later just change the source of the ray to the other side (here $t'_2$. We are only visiting the rays and not the sources itself! 

Next we consider the line $l$ that runs through $s$ and in parallel to the ground level $G$ (which has to be known for the agent for orientation). We can now prolong or shorten the rays toward this line $l$, compare Figure~\ref{fig-terrainred}(ii). If we now also just omit the terrain as in Figure~\ref{fig-terrainred}(iii), we just attain an instance of our pure ray-search problem. This problem seems to become a bit harder to cope with (w.r.t. the competitive ratio) because the shortest path to the rays are now be shortened to a line segment and we do no longer have vision. This is not fully clear at this point because the terrain will also influence or restrict the potential movement of the agent. We will finally see that w.r.t. our strategy the terrain will always be helpful and the worst case is attained without any terrain.  

For the other way round we can really give some sort of reduction. For any ray-search scenario in the half-plane with a set of given allowed rays, we can simply construct a terrain, that also exactly triggers such rays. Compare the transformation back from Figure~\ref{fig-terrainred}(iii) to \ref{fig-terrainred}(iv).

\subsection{Known results for the terrain search}\label{subsect-terrainknown}

\begin{figure}[h]
 \begin{center}
\includegraphics[scale=0.35,page=4]{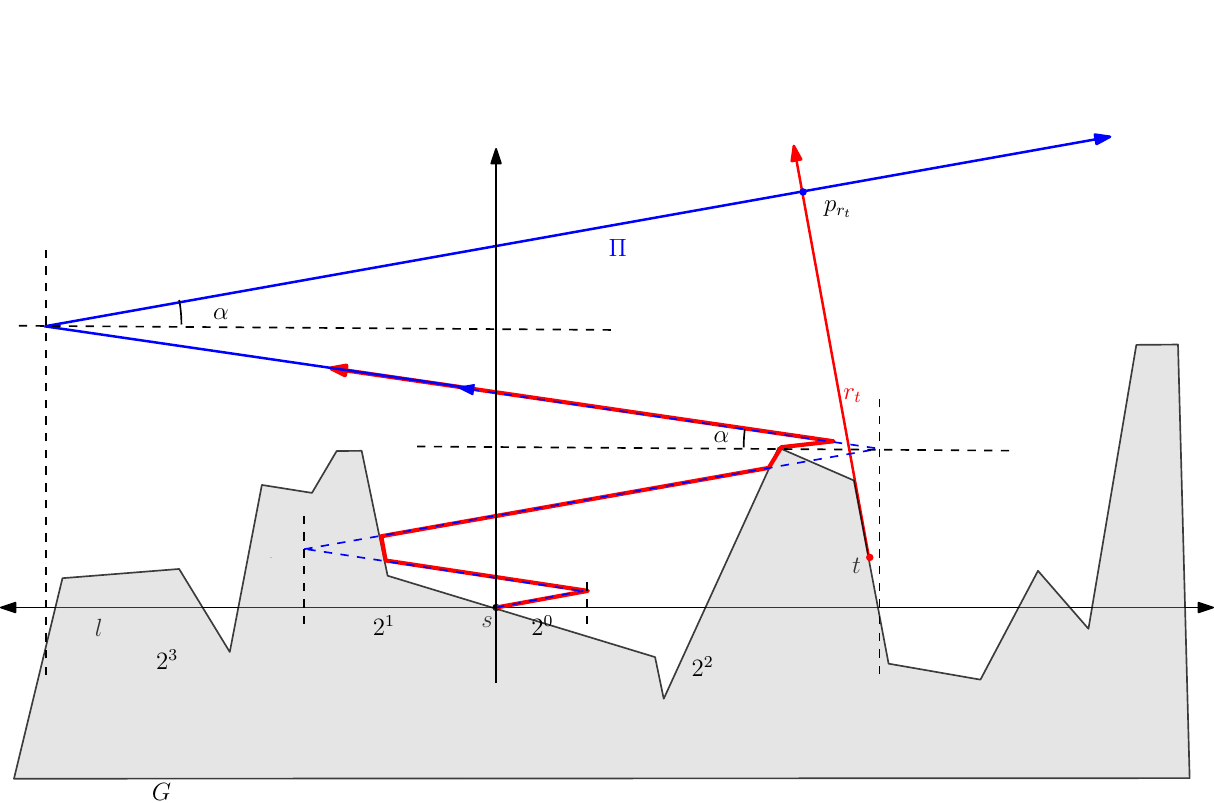}
\caption{Exemplification of the zig-zag strategy plan for the terrain search problem with $x_i=2^i$ and a fixed slope given by angle $\alpha\approx 0.231477$. The return to the other side will always be performed along the fixed pre-computed path. (Left side) The terrain might lead the agent to the next return segment and moves back. (Right side) If the slope direction is free again and the return segment was not met so far, the agent moves into the slope direction for a while.  
}\label{fig-terrain-strat}
\end{center}   
\end{figure}

S.~de~Berg et al. (Latin 2024)~\cite{de2024competitive} presented a strategy for the terrain search problem. They make use of an alternating doubling strategy with $x_i=2^i$ in the $X$-direction and a fixed slope given by 
$\alpha\approx 0.231477$ for any $Y$-movement, see Figure \ref{fig-terrain-strat}. This indeed gives a generalized cow path. 

If the terrain will be hit by this path in a given direction, the agent has to react, starts moving along the terrain and tries to pick up the pre-computed path again, see Figure \ref{fig-terrain-strat}. More precisely, the agent moves in one direction below the next return segment. When the agents hits the terrain, the agent starts following the terrain.  As long as the agent keeps below the return path, she either follows the terrain or she follows the slope direction, if this movement is free. Thus a corresponding next return segment will always finally be found and the path will be picked up again for the movement to the other side. Note that this return path segment has to be free at least up to the starting $X$-coordinate.  

Analytically, in the given paper the doubling distance $x_i=2^i$ was fixed from the very beginning. The slope $\alpha\approx 0.231477363970178$ was attained by considering two cases, one given by a worst case terrain situation and the other given by the above mentioned turning points along the fixed path, similar to the original cow path problem. Both cases were given by vertical rays and a competitive ratio of  $3\sqrt{19/2}\approx 9.24663$ have been guaranteed in general. The authors also show that no strategy can achieve a ratio less than $\sqrt{82}\approx 9.05538$.  

\begin{figure}[h]
 \begin{center}
\includegraphics[scale=0.25]{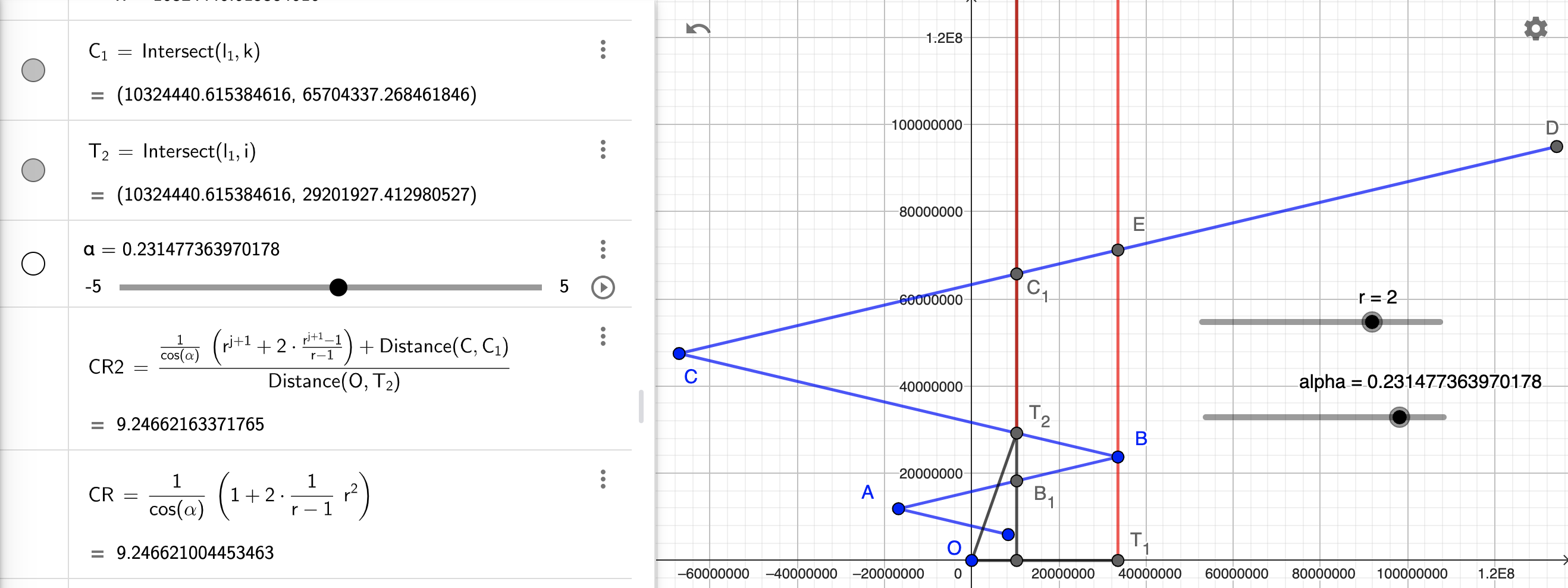}
\caption{The recapitulation and simulation of the worst case situations used by S.~de~Berg et al. (Latin 2024). Both worst case calculations are given for vertical rays and one of those is given without a barrier. The offline shortest path is always a line segment, $OT_1$ or $OT_2$,  respectively. For the terrain case with a single barrier, the path length behind the barrier, $|B_1 B|+|B T_2|$, was taken into calculation although the strategy will omit this path in the given situation. Obviously, there is room for improvement for the terrain case. 
}\label{fig-terrain-strat-geo}
\end{center}   
\end{figure}

For a precise understanding of the two worst case situations we extracted and recomputed both cases with an geogebra-sheet as shown in Figure~\ref{fig-terrain-strat-geo}. Thus it already seems to be clear that the terrain case was a bit too pessimistic because a subpath was unnecessarily taking into consideration. 

\section{Ray search in the half-plane}\label{sect-raysearch}

Due to the above considerations in the following we will proceed in a different way. We first consider the pure ray-search problem without any terrain, we optimize on both parameters, i.e., the doubling factor $r$ and a slope $\alpha_r$ that depends on $r$. The worst case situations are a bit different. Finally, we also adjust the pure strategy for the terrain in a different way. Fortunately, we can easily argue that with the modification we will always profit from the terrain. This also gives more insight into the pure nature of the problem and the analysis appears to be a bit more straightforward. 

\subsection{Design of the strategy and the ratio function}\label{subsect-design}
\begin{figure}[h]
 \begin{center}
\includegraphics[page=5,scale=0.5]{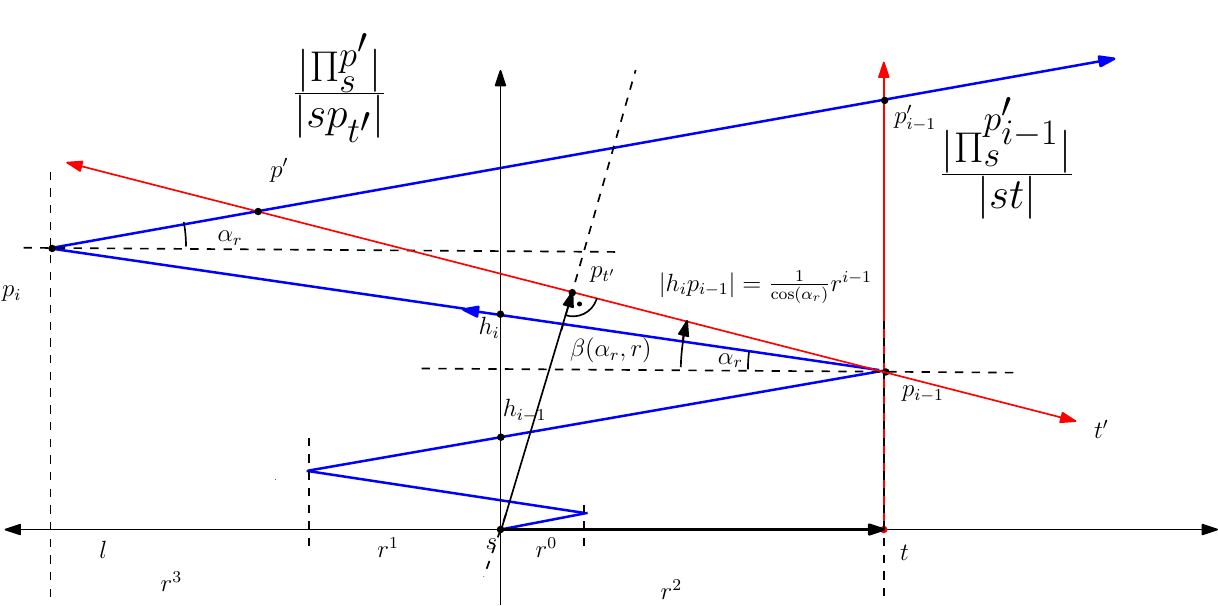}
\caption{Exemplification of the general situation for slope $\alpha_r$ and exponential basis $r$. A local worst case for a ray perpendicular to the baseline is given when the agent slightly misses the ray at point $p_{i-1}$ and detects it at point $p'_{i-1}$. The shortest path to the ray has length $|st|$ along the baseline. In general we consider a local worst case ray with slope $\beta(\alpha_r,r)$ that passes through $p_{i-1}$. The shortest path to this ray is given by the segment $sp_{t'}$ perpendicular to the ray. The agent slightly misses the ray at point $p_{i-1}$ and detects it at point $p'$ after turning back from $p_i$. 
}\label{fig-strategy}
\end{center}   
\end{figure}

We design a cow-path like strategy by using parameters $x_i=r^i$ for the alternating $X$-distances for $i=0,1,2,\ldots$ and $\alpha_r$ for the overall slope,  see Figure~\ref{fig-strategy}. For optimization the slope depends on $r$. 
Let $p_i$ denote the turning points and let $h_i$ denote the heights of the successive visits of the $Y$-axis.

 Figure~\ref{fig-strategy} describes the start and also the general situation. 
A local worst case of the search path $\Pi$ that is similar to the original cow-path is given by a vertical ray $t$ to the left or to the right. The ray $t$ will be slightly missed at a turning point, say $p_{i-1}$, and will be detected at $p'_{i-1}$ in the next turn. The offline shortest path is given by the absolute value of the $X$-coordinate of $p_{i-1}$.  A more general local worst case occurs when a tangent $t'$  to the turning point $p_{i-1}$ is slightly missed and the corresponding ray will be hit at $p'$ after the agent starts the return path on the other side at $p_i$. This worst case depends on the angle $\beta(\alpha_r,r)$ larger than $\alpha_r$ that represents the slope of the corresponding ray $t'$. The agent competes against the shortest path to such a ray $t'$. More precisely, a line segment $Op_{t'}$ perpendicular to $t'$ with $p_{t'}\in t'$.  Obviously, $t'$ and the corresponding path length represent the generic local worst case because $\beta(\alpha_r,r)=\pi/2$ represents the first mentioned worst case where $t'=t$. 

\begin{figure}[h]
 \begin{center}
\includegraphics[page=1,scale=0.5]{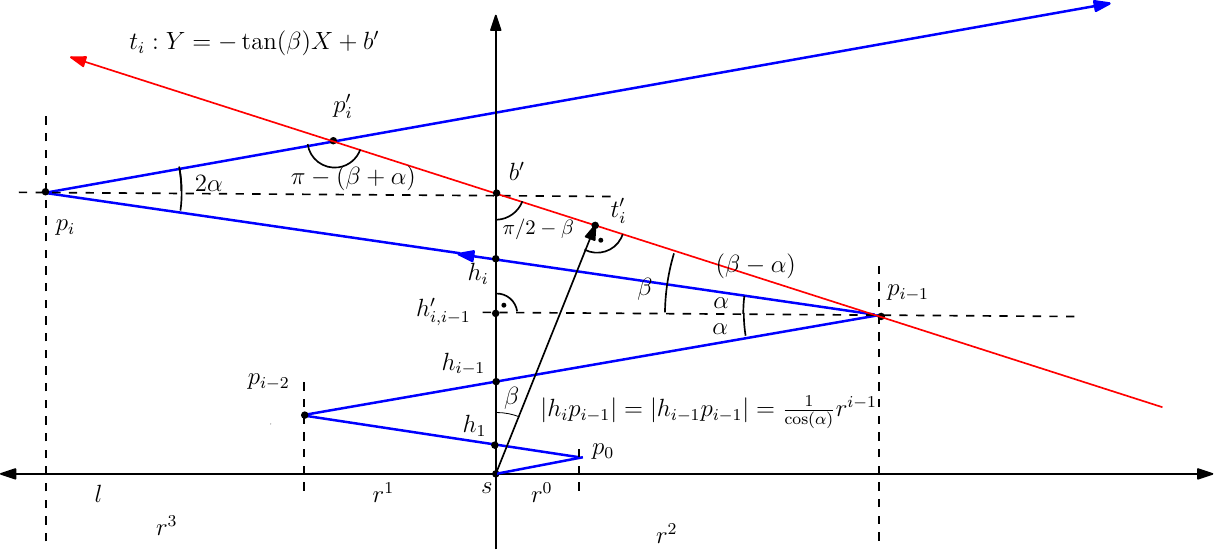}
\caption{The math for the general local worst case subsumes all situations. The ray is part of the line $t_i: Y=-\tan(\beta)X + b'$.  Inserting the known point $p_{i-1}$ gives $b'$. The triangles $s\,t'_i\,b'$ and $p_{i-1}\,h'_{i,i-1}\,b'$ are congruent. Thus we compute $s t'_{i}$ by the law of sine. Similarily the length $|p_i p'_i|$ is computed by the law of sine in the triangle $p_{i-1}\,p'_i\,p_i$. 
}\label{fig-strategy-formula}
\end{center}   
\end{figure}

For convenience we use the notation $\alpha$ and $\beta$ instead of $\alpha_r$  and $\beta_r$, thus implicitly fixing the angles for a fixed $r$. Let $t_i$ denote the above mentioned tangent of the generic local worst case situation  at a turning point $p_{i-1}$ as shown in Figure~\ref{fig-strategy-formula}. Note that precisely the same situation can occur on the right hand side at a turning point $p_{i-2}$ and in any such iteration of the strategy. Asymptotically, there will be no difference. The calculation of the ratio will be the same. We make use of standard trigonometry. 

We have $h_0:=0$ and $h_i=\sum_{j=0}^{i-1} 2 \tan(\alpha)r^{j}=2\tan(\alpha)\frac{r^{i}-1}{r-1}$  for $i=1,2,3,\ldots$ The turning points are given by 
$$p_{i-1}=\left((-1)^{i-1}\cdot r^{i-1},2\tan(\alpha)\left(\frac{r^{i}-1}{r-1}-\frac{1}{2}r^{i-1}\right)\right)$$ for $i=1,2,\ldots$ assuming that we start the move to the right. By symmetry it is sufficient to consider the ratio calculation for the right side, see Figure~\ref{fig-strategy-formula}. For calculating the tangent $t_i$ at $p_{i-1}$ (on the right) with angle $\beta$ we consider

$$t_i: Y= -\tan(\beta)\cdot X + b $$

\noindent
and insert $p_{i-1}$ for calculating $b$. Let $b'$ be the intersection point of $t_i$ with the $Y$-axis. So we have $b=|s b'|=\tan(\beta)r^{i-1}+2\tan(\alpha)\left(\frac{r^{i}-1}{r-1}-\frac{1}{2}r^{i-1}\right)$. Let $t'_i$ be the point on $t_i$ such that $s t'_i$ is the shortest path from $s$ to $t_i$. 
By applying the law of sine to the corresponding (congruent) triangles (compare Figure~\ref{fig-strategy-formula}) we have $|s t'_i|=\sin(\pi/2-\beta)\cdot b=\cos(\beta)\cdot b$, the offline shortest path length to $t_i$. Altogether the shortest path to $t_i$ is 

\begin{equation}\label{equ-shortestpath}
|s t'_i|=\cos(\beta)\cdot \left(\tan(\beta)r^{i-1}+2\tan(\alpha)\left(\frac{r^{i}-1}{r-1}-\frac{1}{2}r^{i-1}\right)\right)\,.
\end{equation}
Next we compute the overall path length of $\Pi_{s}^{p'_i}$ until the ray $t_i$ is detected at $p'_i$. One forth and back movement to a turning point $p_j$ is given by $|h_{j-1}p_j|+|p_j h_j|=\frac{2}{\cos(\alpha)}r^j$. For the path from $p_i$ to $p'_i$ we consider the triangle $p_i$, $p_{i-1}$ and $p_i'$, see Figure~\ref{fig-strategy-formula}. We have $|p_{i-1}p_i|= \frac{1}{\cos(\alpha)}\left(r^{i}+r^{i-1}\right)$ and by the law of sine and the corresponding angles we calculate 
$\frac{|p_i p'_i|}{\sin(\beta-\alpha)}=\frac{|p_{i-1}p_i|}{\sin(\beta+\alpha)}$. Thus the overall path length $|\Pi_{s}^{p'_i}|$ is given by 

 $$|\Pi_{s}^{p'_i}|=\frac{2}{\cos(\alpha)}\left(\sum_{j=0}^{i-1}  r^{j}\right) + \frac{1}{\cos(\alpha)}r^{i}+ \frac{\sin(\beta-\alpha)}{\sin(\beta+\alpha)}\cdot\frac{1}{\cos(\alpha)}\left(r^{i}+r^{i-1}\right)$$

Now we can calculate the (asymptotic) ratio for given $r$, $\alpha$ and $\beta$. Finally, for fixed $r$ and $\alpha$ we can search for the worst case $\beta\in [\alpha,\pi/2]$ which gives the competitive performance of a strategy given by $r$ and $\alpha$.  
We have 
\begin{equation}\label{equ-ratio}
C(r,\alpha,\beta):=\frac{\frac{2}{\cos(\alpha)}\left(\sum_{j=0}^{i-1}  r^{j}\right) + \frac{1}{\cos(\alpha)}r^{i}+ \frac{\sin(\beta-\alpha)}{\sin(\beta+\alpha)}\cdot\frac{1}{\cos(\alpha)}\left(r^{i}+r^{i-1}\right)}{\cos(\beta)\cdot \left(\tan(\beta)r^{i-1}+2\tan(\alpha)\left(\frac{r^{i}-1}{r-1}-\frac{1}{2}r^{i-1}\right)\right)}\,.
\end{equation}

The ratio $C(r,\alpha,\beta)$ in Formula (\ref{equ-ratio}) can be simplified by resolving the geometric sum, multiplication of enumerator and denominator by $\frac{\cos(\alpha)}{r^{i-1}}$, respectively and by taking into account that $\frac{-1}{r^{i-1}}$ will become arbitrarily small for any $r>1$. Thus we (asymptotically) have

\begin{equation}\label{equ-ratio-true}
C(r,\alpha,\beta)=\frac{2\left(\frac{r}{r-1}+\frac{1}{2} r\right)+\frac{\sin(\beta-\alpha)}{\sin(\beta+\alpha)}\left(r+1\right)} {2\cos(\beta)\sin(\alpha)\left(\frac{r}{r-1}-\frac{1}{2}\right)+\cos(\alpha)\sin(\beta) }\,.
\end{equation}

The Formula (\ref{equ-ratio-true}) is well-grounded. For $\beta=\pi/2$ it simplifies to 
\begin{equation}\label{equ-ratio-boundary}
C(r,\alpha,\pi/2)=\frac{2}{\cos(\alpha)}\left(\frac{r^2}{r-1}+\frac{1}{2}\right)
\end{equation} 
which is the worst case for the turning-points, see Figure \ref{fig-strategy}. 
By simple analysis the optimal (minimizing $C(r,\alpha,\pi/2)$ over $r$ for all $\alpha$) is given by $r=2$ and results in $\frac{9}{\cos(\alpha)}$ which gives a ratio of $9$ for the original cow-path problem with $\alpha=0$. 

We can further simplify the ratio formula (\ref{equ-ratio-true}) by simple (i.e., trigonometrical) identities and obtain 
\begin{equation}\label{equ-ratio-true-simple}
C(r,\alpha,\beta)=(r+1) \frac{\frac{r}{r-1} + \frac{\sin(\beta-\alpha)}{\sin(\beta+\alpha)}}{2\cos(\beta)\sin(\alpha)\frac{r}{r-1}+\sin(\beta-\alpha)}\,.
\end{equation}

\begin{equation}\label{equ-ratio-true-simple-second}
C(r,\alpha,\beta)=\frac{r+1}{\cos(\alpha)}\;\;
 \frac{ \frac{r}{r-1} + \frac{\sin(\beta-\alpha)}{\sin(\beta+\alpha)}}{\sin(\beta)+\cos(\beta)\tan(\alpha)\frac{r+1}{r-1}}\,.
\end{equation}

\subsection{Upper bound and analysis of $C(r,\alpha,\beta)$}\label{Sect-subsect-minimize}

We will discuss the optimization of $C(r,\alpha,\beta)$ and make use of MatLab as a standard computer algebra system. Fully analytical optimization seems to be a very tedious task. We have to find the optimal pair $(r,\alpha)$ for $r>1$ and $\alpha\in(0,\pi/2)$ such that $\sup_{\beta\in[\alpha,\pi/2]} C(r,\alpha,\beta)$ is minimal. More formally, the best ratio for a cow-path like strategy is given by

\begin{equation}\label{equ-optformula}
\inf_{(r,\alpha) }\left(\sup_{\beta\in[\alpha,\pi/2]} C(r,\alpha,\beta)\right) \mbox{ for } r>1 \mbox{ and }  \alpha\in(0,\pi/2] \,.
\end{equation}

Obviously, this is a multidimensional optimization problem and not so easy to handle and to solve. 
But we can make use of the advantage that  at first place we are looking for an upper bound rather than providing an overall theoretical optimization of the problem in general. 
For a given fixed cow-path strategy given by a pair a fixed pair $(r,\alpha)$ the optimization over $\beta$ can be finally done analytically (and if necessary numerically). Thus, we finally found our resulting bound of $9.1273$ for all  
$\beta\in[\alpha,\pi/2]$ for the fixed pair $(r,\alpha)$ with $r=1.978624821$ and $\alpha=0.166547577$. At the end we will argue that this choice is well-justified and actually gives no more room for improvements. 

\begin{figure}[h]
 \begin{center}
\includegraphics[scale=0.15]{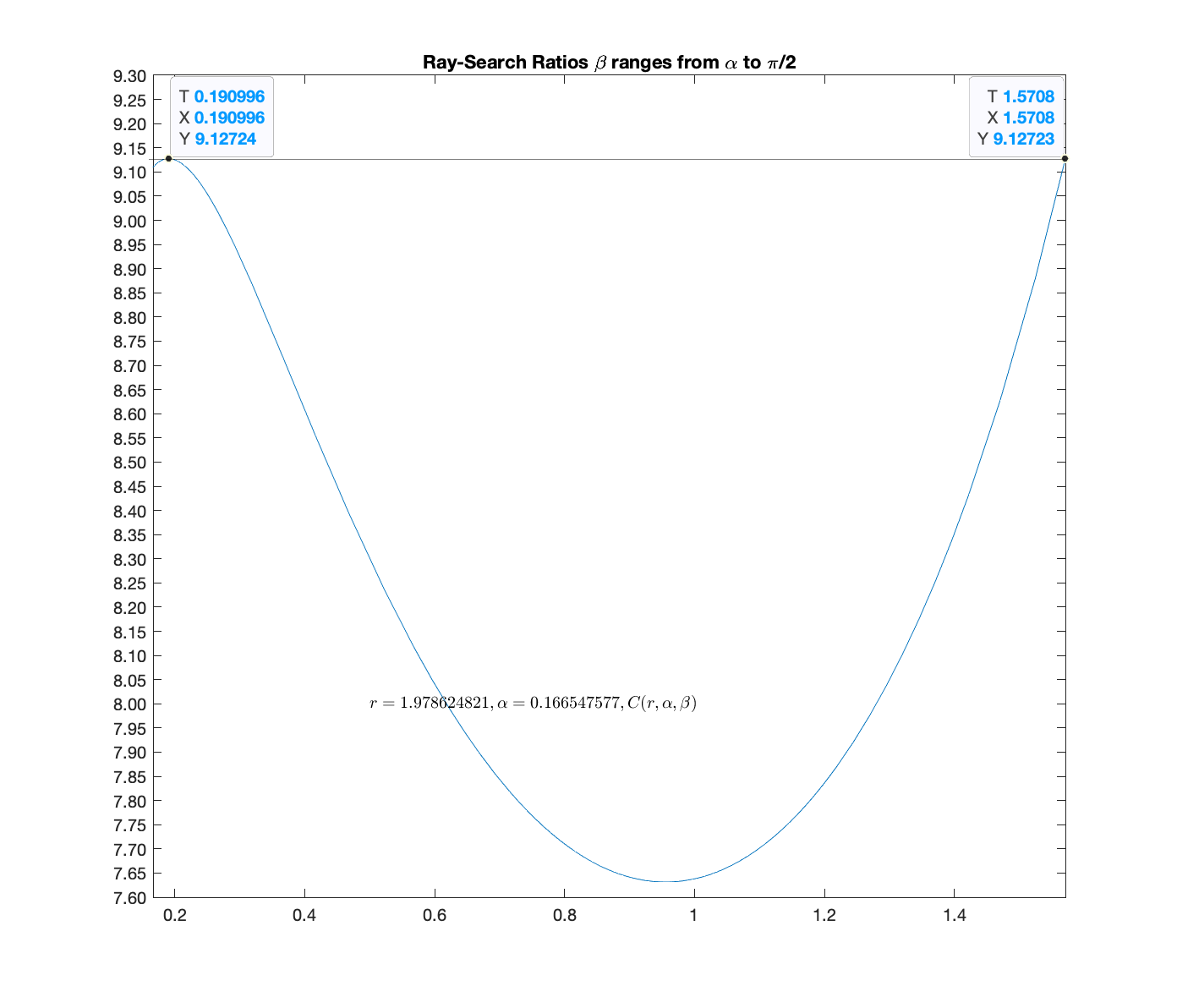}
\caption{ The nature of the general optimization. We have to find pairs $(r,\alpha)$ such that the local maximum $C(r,\alpha,\beta_{\max})$ for some $\beta_{\max}$ shortly after increasing $\beta\in [\alpha,\pi/2]$ and the boundary maximium $C(r,\alpha,\pi/2)$ coincidence. Using $(r,\alpha)$ with $r=1.978624821$ and $\alpha=0.166547577$ gives $C(r,\alpha,\beta)<9.12725$. The value is roughly (say, even slightly smaller) attained by 
$C(r,\alpha,\beta_{\max})$ with $\beta_{\max}\approx 0.190681180073897$ and also by $C(r,\alpha,\pi/2)$.
}\label{fig-final-plot}
\end{center}   
\end{figure}

 Figure \ref{fig-final-plot} shows a plot of this $C(r,\alpha,\beta)$ for $\beta\in[\alpha,\pi/2]$ and also the nature of this bound. The given strategy values balance out two worst case situation, a unique local maximum for $\beta\approx 0.190681180073897$ (computed numerically) and the boundary maximum for $\beta=\pi/2$ are  (up to precision) the same and $C(r,\alpha,\beta)<9.12725$ is guaranteed for all $\beta\in[\alpha,\pi/2]$. So this strategy guarantees the given ratio and the computation of this local and global maximum in a given range can be done numerically and with high precision by MatLab. 
 
We will now briefly discuss how we find these values and why we can finally analytically argue that they optimize Formula (\ref{equ-optformula}) in general (up to precision).

\subsubsection{Interval restriction for $\alpha$ and $r$:}  Here, we already try to argument goal-oriented in order to reasonably restrict the intervals for $\alpha$ and also for $r$. This will be helpful for the analytic interpretation afterwards. 
First, goal-oriented means that we make use of a cow-path like strategy and we would like to keep  strictly below the above mentioned final goal given by $9.12725$, which is our intermediate result so far. For convenience we say that we would like to design a cow-path like strategy better than a ratio of $9.1273$. 

Note that $r=2$ is always (for any $\alpha\in [0,\pi/2)$) the unique local minimum of formula (\ref{equ-ratio-boundary}) and the ratio locally increases for smaller and larger $r$. More precisely by simple analytic means for $r=2$ the function $f(r)=\frac{2}{\cos(\alpha)}\left(\frac{r^2}{r-1}+\frac{1}{2}\right)$ has a unique minimum $f(2)=\frac{9}{\cos(\alpha)}$ for $r>1$ and fixed $\alpha\in [0,\pi/2)$

\begin{figure}[h]
 \begin{center}
\includegraphics[scale=0.5]{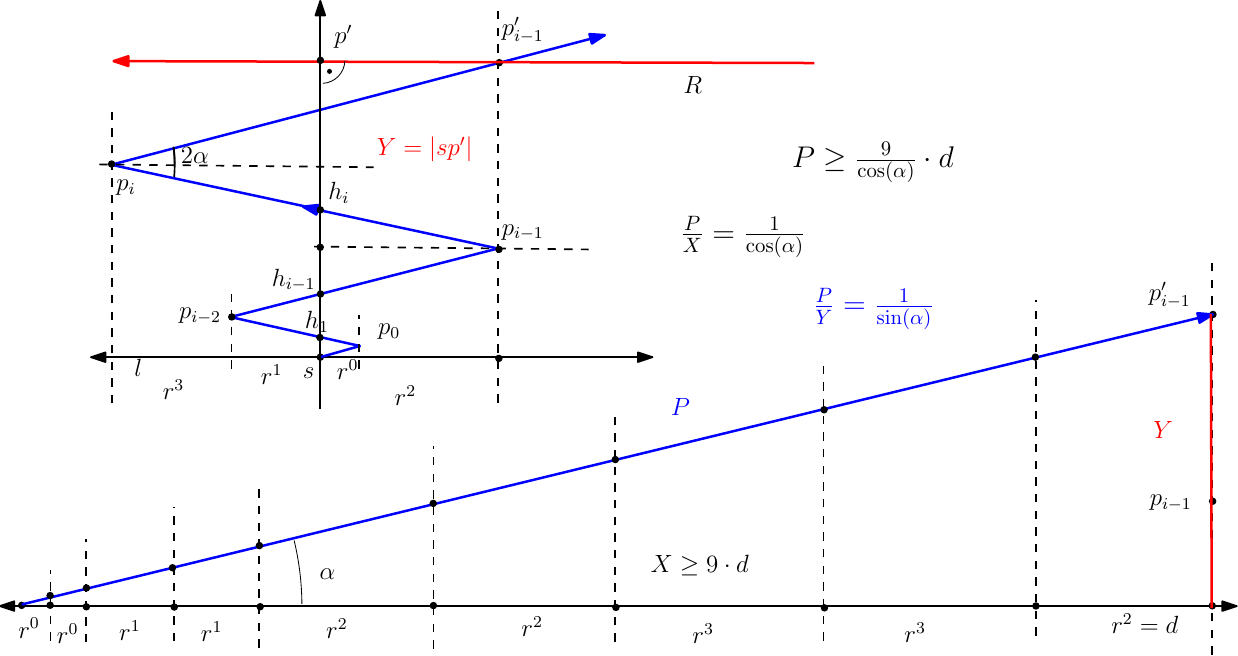}
\caption{For an upper bound on $\alpha$ we unfold the full path of length $P$ up to height $Y$ and consider the local worst case for a horizontal ray at distance $Y$. We have $\frac{P}{Y}=\frac{1}{\sin(\alpha)}$. Note that due to the traditional cow path problem the $X$-distance $X$ is larger than $9\cdot d=9 \cdot r^i$. 
}\label{fig-unfold}
\end{center}   
\end{figure}

Thus we can already plug in that $\alpha$ should be bounded by $\cos(\alpha)> \frac{9}{9.1273}$. This already means that $\alpha$ has to be strictly smaller than $0.1673$. We can also give a reasonable  upper bound on $\alpha$ for our special cow path like strategy. If we fully unfold the path with given fixed $\alpha$ we will reach the height, say $Y$, at a turning point $p_{i-1}$ after a path of overall length, say $P$, was used. An almost horizontal ray at $p_{i-1}$ also gives a bound on $\alpha$. The optimal distance to this ray is $Y$. So we can consider a right-angle triangle with $P$, $Y$ and $\alpha$ where $\frac{P}{Y}=\frac{1}{\sin(\alpha)}<9.1273$ has to hold, see Figure \ref{fig-unfold}. This gives $\alpha>0.109781$. 

On the other hand similar to $\beta=\pi/2$ also $\beta=\alpha$ is a boundary case, the target ray will be detected closely after the next turn. This results in a ratio $C(r,\alpha,\alpha)=\frac{r+1}{\sin(2\alpha)}$. From $\alpha<0.1673$ and the goal $C(r,\alpha,\alpha)<9.1273$ we therefore conclude $r<1.9885$.  Furthermore again for the boundary case $\beta=\pi/2$ and the above lower bound of $\alpha>0.109781$ we conclude 
$r>1.825$. This is shown (again) by the ratio formula (\ref{equ-ratio-boundary}) and the inequality $C(r,\pi/2,0.109781)<9.1273$. Note that as mentioned before $r=2$ is always (for any $\alpha$) the unique local minimum of formula (\ref{equ-ratio-boundary}) and the ratio locally increases for smaller and larger $r$. In turn with this new lower bound on $r$ we can further limit the range of $\alpha$ again. $C(1.825,\alpha,\alpha)=\frac{1.825+1}{\sin(2\alpha)}<9.1273$ gives $\alpha>0.156$. 
And again using formula (\ref{equ-ratio-boundary}) with $\alpha=0.156$ and 
 $C(r,\pi/2,0.156)<9.1273$ results in $r>1.913$.

The overall conclusion is as follows. If we would like to beat a ratio of $9.1273$ with the given strategy design we require $r\in (1.913,1.9885)$ and  $\alpha\in(0.156,0.1673)$ which is an overall reasonable range for numerical optimization, analytic interpretation and reasonable plots.

We can have a closer direct look at the general ratio formula (\ref{equ-ratio-true-simple}) for  any fixed pair $(r,\alpha)$ with $r\in (1.913,1.9885)$ and $\alpha\in(0.156,0.1673)$ and growing $\beta\in[\alpha,\pi/2]$. 
In general for such reasonable values of $r$ and $\alpha$  and for increasing $\beta\in[\alpha,\pi/2]$ the ratio $C(r,\alpha,\beta)$ represents a continuous and differentiable function in this range and behaves in the following way.  First $C(r,\alpha,\beta)$ will increase since the denominator grows slower than the enumerator.  
This will actually happen up to a local maximum. Then the ratio will first decrease to a local minimum and finally --the distance to the ray at the end really shrinks and the path length still increases-- the ratio increases again to a boundary maximum for $\beta=\pi/2$. 
Finally (below) by having a look at the first differentiate of formula~(\ref{equ-ratio-true-simple}) w.r.t. $\beta$ we will analytically argue that this behaviour is given.  

\begin{figure}[h]
\begin{center}
  \begin{minipage}[b]{.44\linewidth} 
  \includegraphics[scale=0.13,page=1]{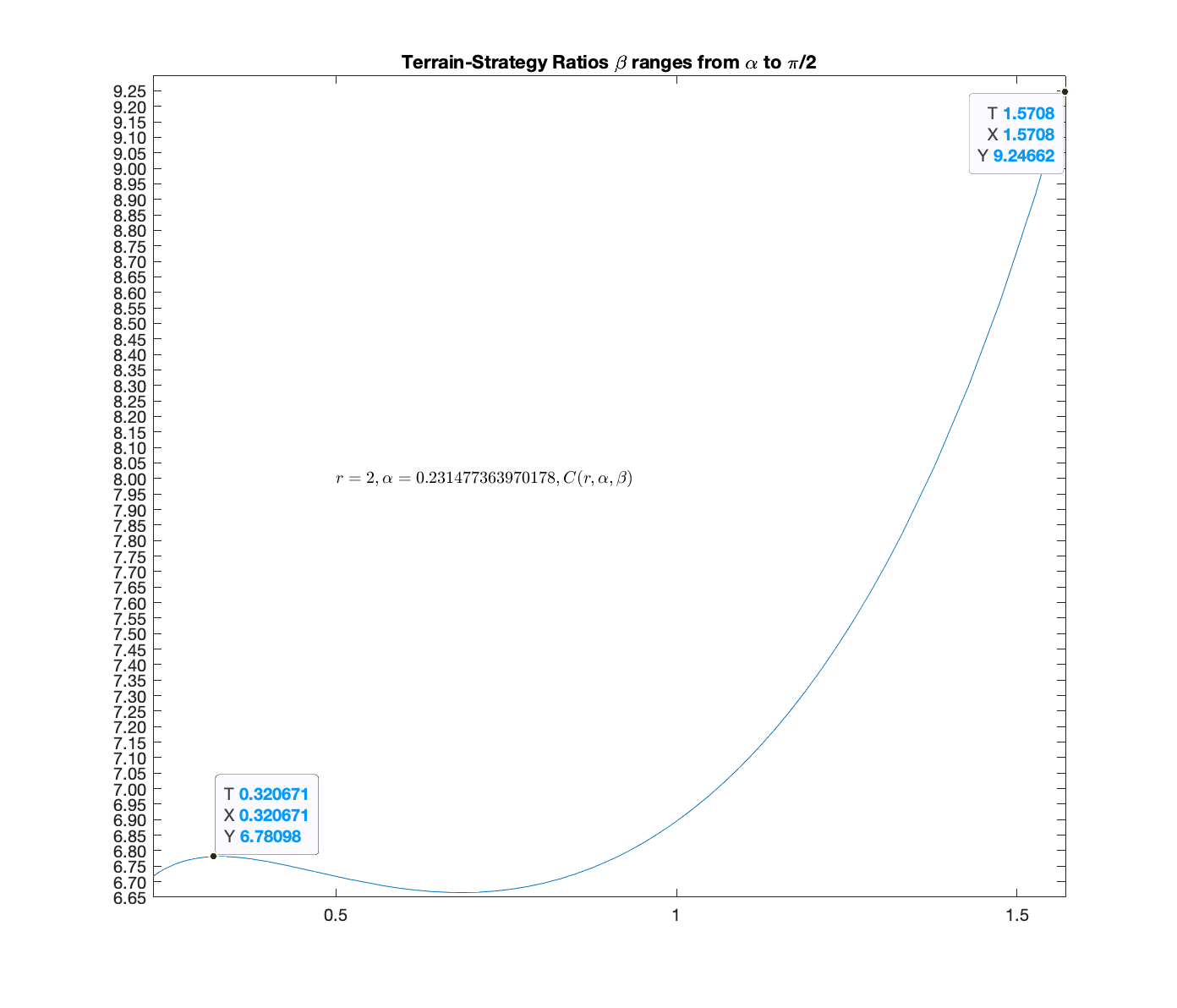}\caption{An example for the case that the boundary maximum $\beta=\pi/2$ subsumes all ratios $C(r,\alpha,\beta)$ with $r=2$ and $\alpha=0.23147\ldots$ taken from \cite{de2024competitive}. }\label{fig-example-plot-terrain}
   \end{minipage}
   \hspace{.1\linewidth}
   \begin{minipage}[b]{.44\linewidth} 
      \includegraphics[scale=0.18,page=2]{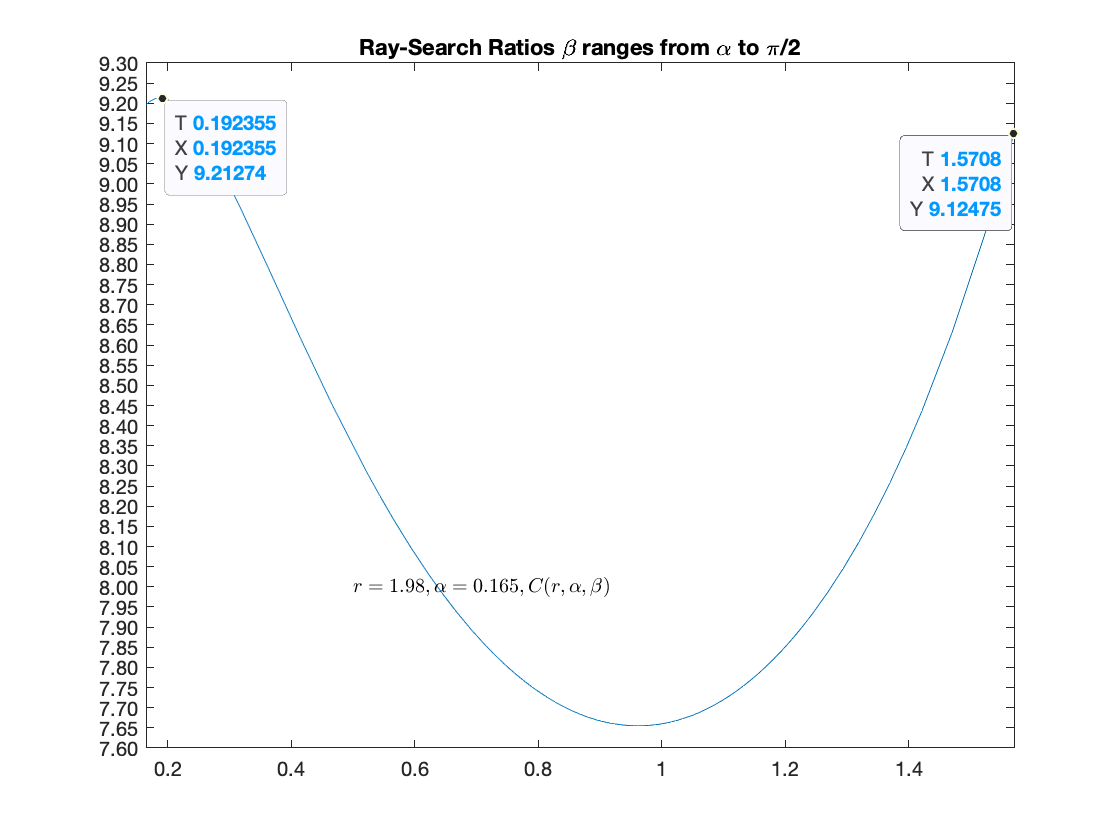}\caption{An example for the case that a local maximum at $\beta\approx 0.192356$ subsumes all ratios $C(r,\alpha,\beta)$ with $r=1.98$ and $\alpha=0.165\ldots$ }\label{fig-example-plot-local}
   \end{minipage}
  \end{center} 
\end{figure}

As an example, if we just use the Terrain-Search strategy values (from \cite{de2024competitive}) for  the pure ray-search strategy and the corresponding plot of $C(r,\alpha,\beta)$ for $r=2$ and $\alpha=0.23147\ldots$ shows that the boundary maximum for $\beta=\pi/2$ subsumes all other ratios, thus guaranteeing the given ratio $9.24663$, see Figure \ref{fig-example-plot-terrain}. This was already our starting conjecture, this is not the best choice. On the other hand for values
$r=1.98$ and $\alpha=0.165$  the given strategy attains a worst case ratio $9.21274$ at the local maximum for some $\beta\approx 0.192355$ closely after changing the cow-path direction, see Figure \ref{fig-example-plot-local}. It seems that we indeed have to balance these two worst case situations out, the local maximum and the boundary maximum should be the same. 

\begin{figure}[!h]
\begin{center}
      \includegraphics[scale=0.4,page=1]{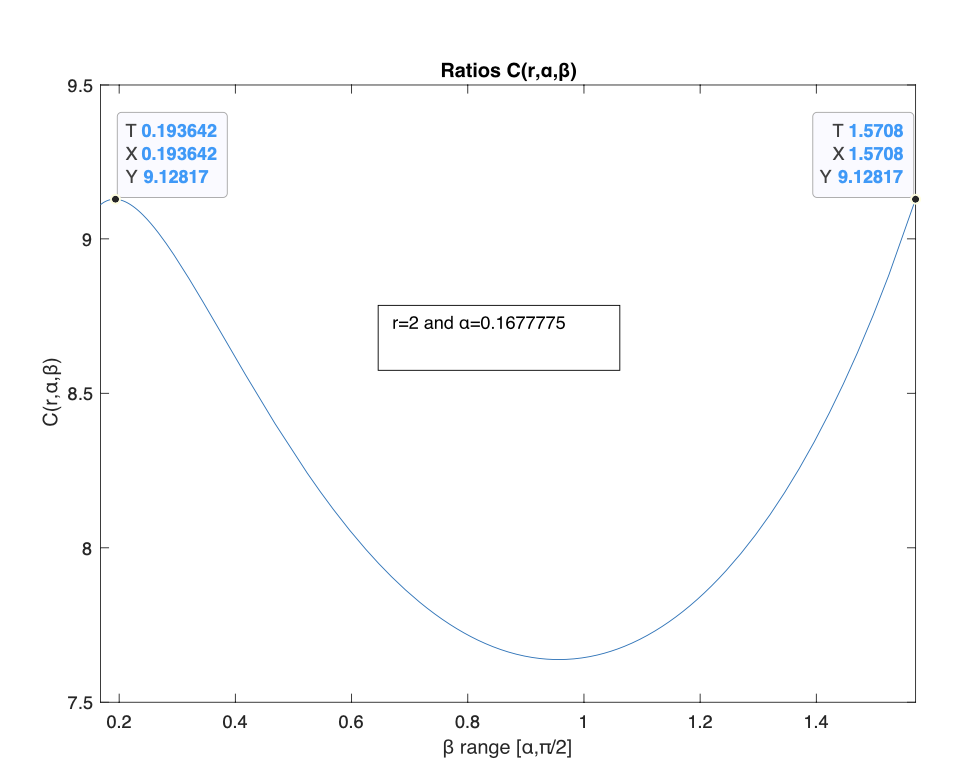}\end{center}
      \caption{Balancing out the worst case ratios (boundary maximum for $\beta=\pi/2$ and local maximum for $\beta\approx 0.19364$) for $r=2$ already results in a very good strategy. For $\alpha\approx 0.1677775$ the strategy guarantees a ratio $C(r,\alpha,\beta)$ less than $9.1282$.}\label{fig-example-plot-balance2}

\end{figure}

We will give some further convincing arguments that the choice of our pair above is almost (apart from numerical precision) optimal among all such strategies. 
The above upper bound on $\alpha$ already shows that we have to keep below $0.1673$ but we also have to adjust $r$. Balancing out the ratio for $r=2$ results in $\alpha\approx 0.167777$ and guarantees already a reasonable ratio less than $9.1282$, see Figure \ref{fig-example-plot-balance2}.  Therefore the only way to keep below this ratio $9.1282$ is to also decrease $r$ slightly, \emph{and} additionally always balancing out the two worst case ratios by adjusting $\alpha$. Thus we numerically found the corresponding best pair $(r,\alpha)$ by making use of MatLab and alternatively also by an interval arithmetic program written by Florian Gans as part of a LAB project.  

\subsubsection{Analytical argumentation:}  
Note that it does not make sense to balance out the two boundary cases $\beta=\alpha$ and $\beta=\pi/2$. Finally, this also means that the above interchanging arguments of minimizing the ranges for $\alpha$ and $r$ will come to an end! As mentioned above for any reasonable pair $(r,\alpha)$ the value of the ratio  (\ref{equ-ratio-true-simple}) will first always increase starting at $\beta=\alpha$. Formally, we can analyse the sign (of the enumerator) of the first differentiate of $C(r,\alpha,\beta)$ w.r.t. $\beta$. We have

$$\begin{array}{l}
C'(r,\alpha,\beta)=(r+1)\times\\[2ex]
  \frac{\left(2\cos(\beta)\sin(\alpha)\frac{r}{r-1}+\sin(\beta-\alpha)\right)\frac{\sin(2\alpha)}{\sin(\beta+\alpha)^2}+\left(2\sin(\beta)\sin(\alpha)\frac{r}{r-1}-\cos(\beta-\alpha)\right)\left( \frac{r}{r-1} + \frac{\sin(\beta-\alpha) }{\sin(\beta+\alpha)}\right)}{\left( 2\cos(\beta)\sin(\alpha)\frac{r}{r-1}+\sin(\beta-\alpha)\right)^2}\,.\\[1ex]
\end{array}$$

Now  from $C'(r,\alpha,\alpha)=\frac{(r+1)}{2\cos(\alpha)^2}>0$ we conclude that in beginning for growing  $\beta\in [\alpha,\pi/2]$ the ratio $C(r,\alpha,\beta)$ grows. 
Additionally, we can make use of the intervals for $r$ and $\alpha$. 
For all such pairs $(r,\alpha)$ we can show that  $C'(r,\alpha,2\alpha)<0$ holds and there is a local maximum of $C(r,\alpha,\beta)$ for $\beta\in[\alpha,2\alpha]$. To show this we just minimize the chances of the enumerator of $C'(r,\alpha,2\alpha)$ to be negative! By taking the smallest or largest $r$ and  the smallest or largest $\alpha$, respectively, if it helps to make  $C'(r,\alpha,2\alpha)$ positive it still remains negative. As already mentioned the ratio finally has to grow, $C'(r,\alpha,\pi/2)>0$ always holds. Therefore there is also a local minimum in $\beta\in[2\alpha,\pi/2]$. More precisely already $C'(r,\alpha,\pi/2-\alpha)>0$ can be shown in the given intervals. 

\begin{figure}[h]
 \begin{center}
\includegraphics[scale=0.15]{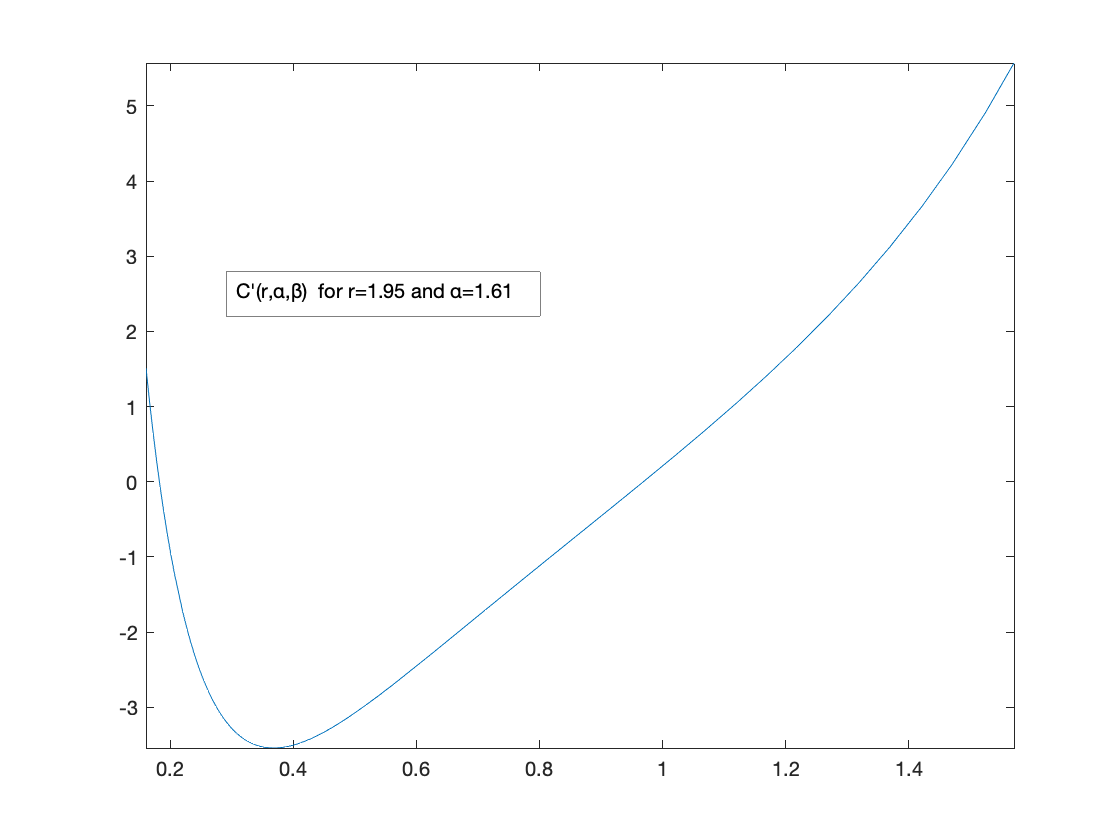}
\caption{ For any reasonable pair $(r,\alpha)$ the differentiate of $C(r,\alpha,\beta)$ in $\beta$ has two zeros that belong to a maximum of $C(r,\alpha,\beta)$ for $\beta \in[\alpha,2\alpha]$ and a minimum of $C(r,\alpha,\beta)$ for $\beta \in[2\alpha,\pi-\alpha]$. Finally there is a boundary maximum at $\beta=\pi/2$. 
}\label{fig-diff-ex}
\end{center}   
\end{figure}

In total we can argue that the function $C'(r,\alpha,\beta)$ for $\beta \in [\alpha,\pi/2]$ and for any pair $(r,\alpha)\in [(1.913,1.9885),(0.156,0.1673)]$ starts with a positive value, has some negative minimum and gets positive again, see Figure \ref{fig-diff-ex} for an example for $r=1.95$ and $\alpha=1.61$. So the differentiate becomes zero twice and in turn $C(r,\alpha,\beta)$ behaves as already mentioned with a local maximum for some $\beta\in[\alpha,2\alpha]$, a local minimum for some $\beta\in[2\alpha,\pi/2-\alpha]$ and a boundary maximum for $\beta=\pi/2$. Therefore it is clear what has to be balanced out! We can guarantee a ratio less than $9.12725$ by using $r=1.978624821$ and $\alpha=0.166547577$.

\subsection{Lower bound constructions}\label{subsect-lower}

For the lower bound on a competitive ratio for any strategy we first argue that any reasonable strategy will monotonically increase the $Y$-coordinates. There is no benefit in moving down. Additionally, any reasonable strategy will successively and alternatingly extend the $X$-direction to the left and to the right as in the traditional cow-path problem. Similar to the traditional cow-path problem there is always a local worst case situation for a vertical ray when the strategy comes back to a previous maximal distance~$d$ on the left or right. At this point also the current highest overall $Y$-coordinate $d_y$ is visited and we also compete against a horizontal ray of distance $d_y$ from the start. 

\begin{figure}[h]
 \begin{center}
\includegraphics[scale=0.5]{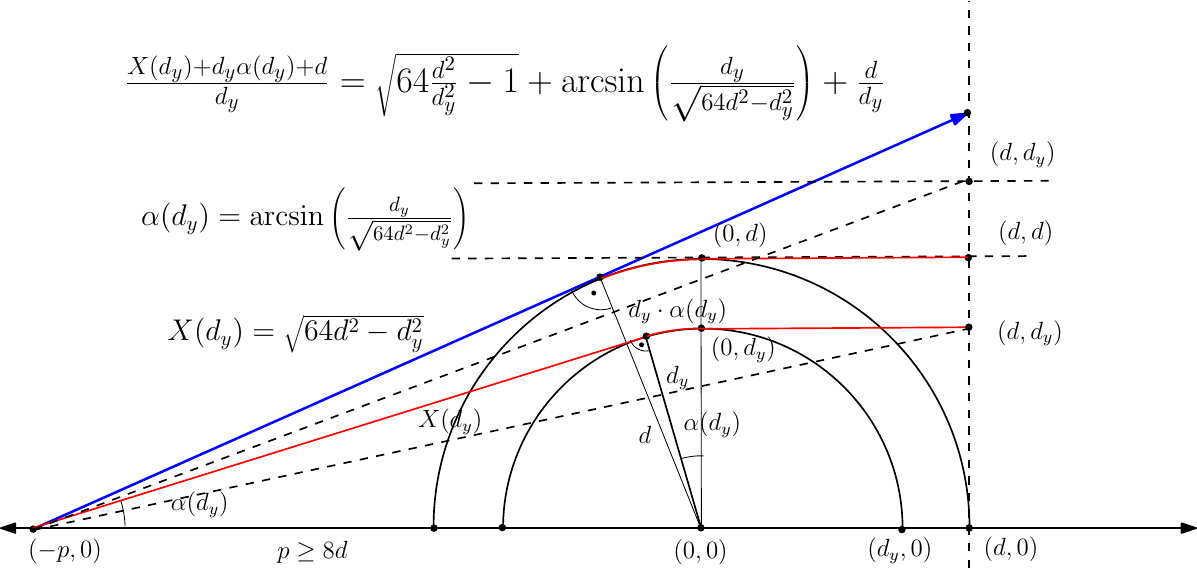}
\caption{We completely unfold and shorten (by using a straight line) the full path of an arbitrary strategy up to a local worst case vertical ray with $X$-direction~$d$. There we hit the point $(d,d_y)$ of maximal height among all points of the strategy.  Since we consider the shortest direct movement to $(d,d_y)$ we can first just consider a triangle $(-p,0),(d,0),(d,d_y)$ and a vertical ray and a horizontal ray both passing $(d,d_y)$ for comparing two ratios. Note that $p\geq 8d$ can be assumed and we make use of $p=8d$ for the lower bound. The minimum of the maximum of both is attained for $d_y=d$. For the improvement of the lower bound we can additionally assume that the unfolded path actually has to keep outside the half-circle of radius $d_y$ around the start in order to compete against all rays with distance $d_y$ to the start. The shortest such circle avoiding path divided by $d_y$ gives a lower bound. For the lower bound we can only guarantee the minimum for all $d_y\leq d$ which is attained again for $d=d_y$. 
}\label{fig-unfold-gen}
\end{center}   
\end{figure}

So we consider an arbitrary strategy ans 
as already used in Figure \ref{fig-unfold} we just unfold its total path and look at such a worst case situation for the $X$-direction at some arbitrary stage of the strategy. 
There has to be such a local worst case point $(d,d_y)$ with overall $X$-distance path length larger than $9\times d$ (which already comes from the regular cow-path problem). For a lower bound on the current path distance we now just consider the simple straight line path to $(d,d_y)$ after the total enfolding of the strategy this path length is smaller than the length of the original strategy. In Figure \ref{fig-unfold-gen} we only exemplify this unfolding and the final path simplification by the triangles $(-p,0),(0,d),(d,d_y')$. 
Note that $p\geq 8 d$ can be assumed. In the following we will set $p=8d$ because this will always minimizes the ratios of the constructed lower bounds. 

We now can calculate two ratios  $\frac{\sqrt{81d^2+(d_y)^2}}{d}$  and  $\frac{\sqrt{81d^2+(d_y)^2}}{d_y}$ for the case of a vertical ray with distance $d$ or a horizontal ray with distance $d_y$, respectively.  Either $d_y\geq d$ or $d\geq d_y$ holds. This  means that either the first or the second ratio is worse. Thus for an overall lower bound in this setting we have to calculate the ratio for $d=d_y$ which minimizes the maximum of the two ratios. This gives a first lower bound of $\sqrt{82}\approx 9.0553851\ldots$ which was already known. We can further improve this lower bound. 

At the point $(d,d_y)$ we have to take care that we not only compete against a horizontal and vertical ray, we have to consider all rays with shortest distance $d$ or $d_y$ from the start, compare Figure~\ref{fig-unfold-gen} again. 

 Competing against all rays means that  the shortest unfolded path to  $(d,d_y)$ has to keep outside the half-circle of radius $d$ or $d_y$ around the start.
A lower bound to such a shortest unfolded path will hit this circle by a tangent, then move along the circle  and then (tangentially) move toward $(d,d_y)$. It is not forbidden that $d_y$ is larger than $d$ but then we just always compete against the vertical ray of distance $d$ and also its corresponding circle, this will result in a larger ratio. 

 \begin{figure}[h]
 \begin{center}
\includegraphics[scale=0.15]{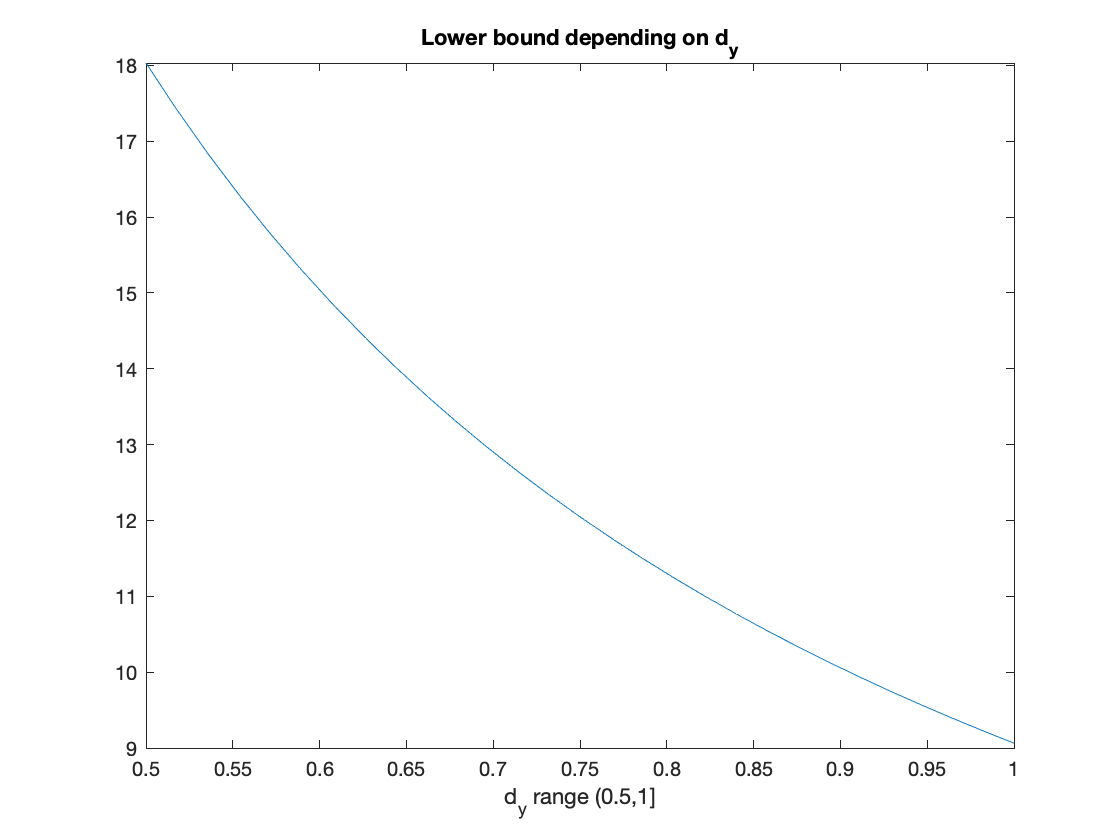}
\caption{The plot of the monotonically decreasing ratio function $R(d_y)$ with $R(1)=9.063577796336\ldots$ as an overall lower bound.
}\label{fig-lb-funktion}
\end{center}   
\end{figure}

Therefore for the calculation of a lower bound for this setting we can simply express the corresponding ratio as a function of $d_y$ for $d_y\in [0,d]$ using a half-circle of radius $d_y$ at the origin. So the path length to the local worst case is at least a tangential movement of distance $X(d_y)=\sqrt{64d^2-d_y^2}$ 
to the circle plus an arc of length $\alpha(d_y)\cdot d_y$  where $\alpha(d_y)=\arcsin\left(\frac{d_y}{\sqrt{64d^2-d_y^2}}\right)$ and a  distance $d$ from $(0,d_y)$ to $(d,d_y)$. The calculations are shown in Figure~\ref{fig-unfold-gen}.  We get a ratio that depends on $d_y\in [0,d]$. Since we consider the ratios we can also scale the situation such that $d=1$ thus calculating the minimum (as a proven lower bound) of 
$$R(d_y):=\sqrt{64\frac{1}{d_y^2}-1}+\frac{1}{d_y}+\arcsin\left(\frac{d_y}{\sqrt{64-d_y^2}}\right)$$
for $d_y\in (0,1]$.  The ratio function $R(d_y)$ is monotonically decreasing (see Figure~\ref{fig-lb-funktion}) and $d_y=1$ thus gives the lower bound of $R(1)=9.063577796336\ldots$ There is no strategy with a better ratio!

\section{Application and adaptation for the terrain problem}

We now would like to apply and adjust the above cow path strategy to the terrain search problem. The first idea that would came into mind is, just taking the zig-zag strategy idea from S.~de~Berg et al. (Latin 2024)~\cite{de2024competitive} and use the adjustments presented in Figure~\ref{fig-terrain-strat}. It turns out that this is not overall helpful, we should make use of the terrain in order to exploit that the shortest paths to current worst case rays will really grow into the $Y$-direction. Therefore in contrary to the previous ideas when we hit the terrain  we will adjust the pure strategy in a different way presented as follows.

Note that we finally will use the parameters $r=1.978624821$ and $\alpha=0.166547577$  from Section~\ref{Sect-subsect-minimize} and in the following we consider this original pure strategy as $\Pi_{\mbox{orig}}$.
We can compute these values by Matlab with very high precision in order to keep below the given ratio. The precise values will be taken into account in the latest moment. As long as possible we will do the analysis symbolically. At the end we will be sure that the calculations are correct and do not depend on numerical problems.

\begin{figure}[h]
 \begin{center}
\includegraphics[scale=0.45,page=4]{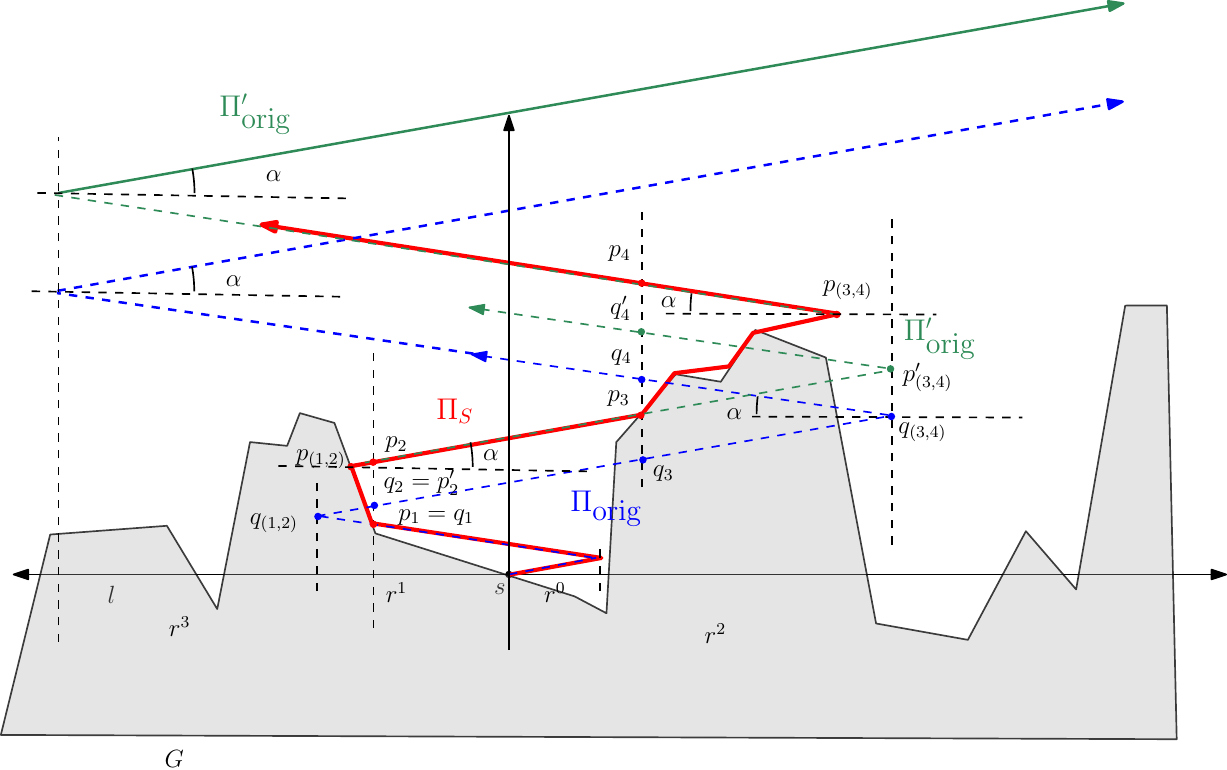}
\caption{Exemplification of the new zig-zag strategy adaption for the terrain search problem with $X$ distances $r^i$ and a fixed slope given by angle $\alpha$. We use the original strategy $\Pi_{\mbox{orig}}$ or its copies  $\Pi'_{\mbox{orig}}$ and in case of collisions with the terrain we use the terrain to climb up. The new strategy $\Pi_S$  attains the same $X$-coordinates in the free space back movement,  compare point $p_{4}$ in comparison to $q_{4}$ on $\Pi_{\mbox{orig}}$ and/or $q'_{4}$ on $\Pi'_{\mbox{orig}}$. Also by construction the overall path length remains the same up to $p_4$ and $q_4$, respectively. The point $p_4$ potentially have a larger $Y$-coordinate than $q_4$. For example when the terrain is visited for the first time on the current intermediate plan $\Pi'_{\mbox{orig}}$ at some point $p_{3}$ we use the budget $|\Pi_{\mbox{orig}}(q_{3},q_{4})|$ (which is also $|\Pi'_{\mbox{orig}}(p_{3},q'_{4})|$) and climb up along the terrain until we have to follow the $\alpha$-direction again or we have to return from a point $p_{(3,4)}$ to $p_{4}$ in order to locally guarantee that
 $|\Pi_{\mbox{orig}}(q_{3},q_{4})|=|\Pi_{S}(p_{3},p_{4})|$.  The return point $p_{(3,4)}$ can be computed during the movement, and we move back into $\alpha$-direction in the free space and reach $p_4$ as intended with overall guarantee 
 $|\Pi_{\mbox{orig}}(s,q_{4})|=|\Pi_{S}(s,p_{4})|$!
}\label{fig-terrain-strat-adapt}
\end{center}   
\end{figure}

In the following let $\Pi_S$ denote our strategy adjusted  according to the terrain.  In the beginning $\Pi_{\mbox{orig}}$ and $\Pi_S$ are just the same and we start moving along $\Pi_{\mbox{orig}}$. 
If we hit the terrain with the (pure) strategy at some point say $q_1$ for the first time we suggest the following adjustment. As an invariant we recently have followed some copy of the original strategy into some direction but may be with larger $Y$-coordinate. Actually starting from $q_1$ after a turn at some point $q_{(1,2)}$ we originally had planned to move back to the same $X$-coordinate to some point $q_2$ on the original strategy, compare Figure~\ref{fig-terrain-strat-adapt}. 

Since such a movement is no longer possible starting at $p_1=q_1$ we leave the original idealistic path and follow the terrain for a while.  The  plan is to finally visit a point $p_2$ with the same $X$-coordinate as $q_2$ and $q_1$ and with larger $Y$-coordinate. And at $p_2$ we also  would like to be back on a free path back into $\alpha$-direction to the other side, and thus starting with a new copy $\Pi'_{\mbox{orig}}(p_2)$ of $\Pi_{\mbox{orig}}(q_2)$ above $q_2$. Furthermore the partial new path $\Pi_{S}(p_1,p_2)$ should have the same path length as $\Pi_{\mbox{orig}}(q_1,q_2)$ before, i.e., we would like to have $|\Pi_{\mbox{orig}}(q_1,q_2)|=|\Pi_{S}(p_1,p_2)|$. To attain these properties starting at $p_1$ we just make use of the budget  $|\Pi_{\mbox{orig}}(q_1,q_2)|$ and start moving along  the terrain. During the movement we can leave the terrain into $\alpha$-direction, when the $\alpha$-direction becomes free again or when we need the remaining budget to return back in $\alpha$-direction to the previously mentioned $X$-coordinate of $q_2$. In the latter case we have just reached a turning point $p_{(1,2)}$. It is obviously possible to always calculate the rest budget and also the return point $p_{(1,2)}$ during the movement appropriately. From this point $p_{(1,2)}$ we move back in  $\alpha$-direction and this path has to be free, see Figure~\ref{fig-terrain-strat-adapt}. Thus we are finally located at $p_2$ vertically above $q_2$ and the overall path length of the adapted strategy from the very beginning has the same path length as the original part, we guarantee $|\Pi_{\mbox{orig}}(s,q_2)|=|\Pi_{S}(s,p_2)|$. 

Now we can make use of the invariant. We will now follow again a copy $\Pi'_{\mbox{orig}}(p_2)$ of the original strategy $\Pi_{\mbox{orig}}(q_2)$ by starting at $p_2$ but vertically above $q_2$ with a larger $Y$-coordinate. This was attained just because part of the original strategy path length was just used on the terrain and we only run upwards a bit! 

We now again idealistically follow a copy $\Pi'_{\mbox{orig}}(p_2)$ of the original strategy until we will hit the terrain on the other side (or may be even later somewhere). In Figure~\ref{fig-terrain-strat-adapt} this happens for example at point $p_3$ and we apply the same rule. The original \emph{new} partial plan was moving forth and back from $p_3$ over the turning point $p'_{(3,4)}$ to $p_4'$ on $\Pi'_{\mbox{orig}}(p_3)$. Since this is not possible any more we follow the terrain again and exploit the budget $|\Pi'_{\mbox{orig}}(p_3,p_4')|$ (which is obviously the same as $|\Pi_{\mbox{orig}}(q_3,q_4)|$ from the overall original strategy). Again we would like to reach  a point $p_4$ with the same $Y$-coordinate as $p_4'$ (and also $q_4$) along $\Pi_S(p_3)$. During the movement we will leave the terrain, when the $\alpha$-direction becomes free again or when we need the remaining budget to return back in $\alpha$-direction from a turning point $p_{(3,4)}$ to the previously mentioned $X$-coordinate of $p_4'$. It is obviously possible to calculate the return point $p_{(3,4)}$ appropriately during this movement taking the remaining budget into account.  From this turning point we move back in  $\alpha$-direction and this path has to be free definitely, see Figure~\ref{fig-terrain-strat-adapt}. We will reach $p_4$ on $\Pi_{S}$. Now $|\Pi_{\mbox{orig}}(s,q_4)|=|\Pi_{S}(s,p_4)|$ holds and $p_4$ lies vertically above $q_4$. 
 
 The general policy is as follows:
Overall the adjusted  strategy $\Pi_S$ attains the same $X$-coordinates in the free space at points $p_{2i}$ in comparison to $q_{2i}$ on $\Pi_{\mbox{orig}}$ and/or $q'_{2i}$ on the current $\Pi'_{\mbox{orig}}$. We move on the current copy $\Pi'_{\mbox{orig}}(p_{2i-2})$ of $\Pi_{\mbox{orig}}(q_{2i-2})$. When the terrain is visited on the current intermediate plan at some point $p_{2i-1}$ we use the budget $|\Pi_{\mbox{orig}}(q_{2i-1},q_{2i})|$ and climb up along the terrain until we have to follow the $\alpha$-direction again or we have to return from $p_{(2_i-1,2i)}$ to $p_{2i}$ in order to always guarantee 
 $|\Pi_{\mbox{orig}}(q_{2i-1},q_{2i})|=|\Pi_{S}(p_{2i-1},p_{2i})|$. From $p_{2i}$ on we use the next copy $\Pi'_{\mbox{orig}}(p_{2i})$ of $\Pi_{\mbox{orig}}(q_{2i})$.

In the following we will show that this simple strategy adaptation for the terrain with the given parameters will always guarantee a competitive ratio for the terrain search problem that is not worse than the original pure strategy. The terrain gives us a real benefit. 

Also note that a situation as in Figure~\ref{fig-terrain-strat-adapt} might be just a starting situation. And may be the terrain does not go up further in $Y$-direction on the boundaries but just keeps above the line $l$ with some small peaks for the simulation of arbitrary rays and the low level terrain goes to infinity on both sides. This means that at the end the starting benefit will be totally subsumed and we are back in the analysis of the pure strategy. In this case the starting benefit is just a negative additive constant and we can actually only guarantee the ratios of the pure strategy!


\subsection{Performance guarantee by geometric arguments}

In the analysis of the pure strategy we balanced out the two worst case situations, one as a local maximum after a return and the other one as a boundary maximum. The advantage of the local adjustment is that after some return point $p_{(2i-1,2i)}$ we visit the same rays  \emph{earlier} in time with shorter path distance. Thus for the local maxima in the analysis before, we obtain better ratios somehow at hand. 
Considering the worst case boundary case our adjustment could mean that the 
 worst case vertical ray might have  smaller $X$-distance to the start than before. But visiting the terrain in between and taking the original worst case into account shows that the ratio will indeed also decrease, the length of a shortest path to the vertical ray increases, the path also moves along the terrain. We will prove these arguments geometrically in the following. 

\subsubsection{Decreasing ratios for the rays:} 
\begin{figure}[h]
 \begin{center}
\includegraphics[scale=0.45,page=4]{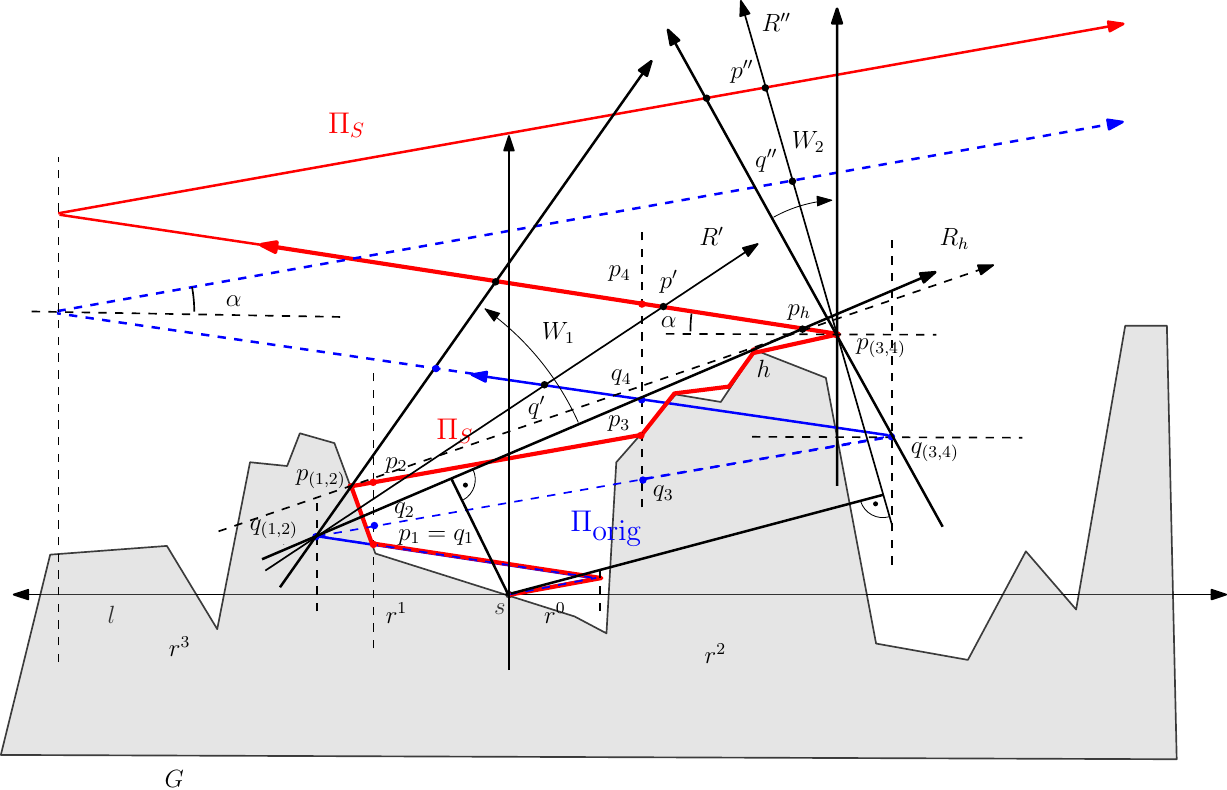}
\caption{Exemplification of the ratio improvement if the strategy hits the terrain. There are local worst case rays after the strategy start returning at $p_{(3,4)}$  To make it locally even worse for the ratios of  $\Pi_S$ we consider the turning point $q_{(1,2)}$ of $\Pi_{\mbox{orig}}$ as the source of the wedge of rays. 
For rays $R'$ in the wedge $W_1$ we can guarantee a ratio not worse than the original strategy, since $|\Pi_{S}(s,p')|\leq|\Pi_{\mbox{orig}}(s,q')|$ holds by construction of the strategy. The wedge $W_1$ ends when $R'$ runs through $p_{(1,2)}$ and $q_{(1,2)}$. In general there is another wedge $W_2$ that has to be considered. The wedge starts with a ray running through two turning points $p_{(2i-1,2i)}$ and $q_{(2i-1,2i)}$ and ends at a vertical ray passing though $p_{(2i-1,2i)}$. The rays $R''$ of wedge $W_2$ rotate around $p_{(2i-1,2i)}$. They will be detected by the next move into the same direction at intersection points $p''$. 
 }\label{fig-terrain-strat-adapt-analysis}
\end{center}   
\end{figure}

For the case without a terrain we had two worst case situations. Now we argue that an adjustment induced by the terrain indeed helps to lower down the ratio locally for all possible rays.  We will indicate two ray intervals $W_1$ and $W_2$ that has to be considered. Those rays will be considered as the local worst case rays for strategy~$S$. We subdivide the rays of interest into two locally fixed bundles $W_1$ and $W_2$ and show that the ratio for $S$ is not larger than the ratio of the original strategy without the terrain. The rays of $W_1$ and $W_2$ are local worst case situations. 

\subsubsection{Decreasing ratios for the rays in $W_1$:} 
One worst case was attained shortly after the return point of the original strategy starting with angle $\beta=\alpha$. 
We now argue that the \emph{new} return points lead to better ratios for a reasonable set of worst case rays detected by the original strategy. The adjusted strategy will first return a bit earlier but on a higher $Y$-level. 
Figure~\ref{fig-terrain-strat-adapt-analysis} locally exemplifies the overall situation.
Let us assume that  the movement along the terrain started at $p_{2i-1}$ and moves to $p_{2i}$. The movement along the terrain ends at a last point $h=(h_x,h_y)$ of maximal height. After that the strategy might still move into the free space into direction $\alpha$ again but it will always return at a new return point $p_{(2i-1,2i)}$ for moving to $p_{2i}$.  The rays of interest (that not have been visited yet) have to keep beyond this point $h$ and also beyond the current strategy part $\Pi_{\mbox{orig}}(q_{(2i-3,2i-2)})$. By construction this part always cannot run above $\Pi_S(p_{(2i-3,2i-2)})$. Therefore, to make it locally worse for  the ratios of  $\Pi_S$, we consider the original turning point $q_{(2i-3,2i-2)}$ of $\Pi_{\mbox{orig}}$ on the other side and consider the ray 
$R_h$ moving through $h$ and $q_{(2i-3,2i-2)}$. This ray has to hit $\Pi_S$ at $p_h$ after the return from $p_{(2i-1,2i)}$. Compared to the ray passing through $h$ and $p_{(2i-3,2i-2)}$ (the previous returning point of $\Pi_S$ not below  $q_{(2i-3,2i-2)}$) the ray $R_h$ has a  distance to the start that is not larger. 

Therefore $R_h$ can be regarded as a starting worst case ray after the return for $\Pi_S$. This ray hits the original strategy $\Pi_{\mbox{orig}}$ later in time than 
$\Pi_S$ at $p_h$. Therefore the ratio for $\Pi_{\mbox{orig}}$ is locally larger when the first new ray is detected by the strategy after the return.

How long can we guarantee this behaviour? We detect a wedge $W_1$ that starts with $R_h$, compare Figure~\ref{fig-terrain-strat-adapt-analysis} and consider a ray $R'$ that runs through $q_{(2i-3,2i-2)}$ of $\Pi_{\mbox{orig}}$ and $p'$ on $\Pi_S(p_h)$ after $p_h$.
Such a ray hits $\Pi_{\mbox{orig}}$ earlier in time at $q'$, more precisely $|\Pi_{\mbox{orig}}(s,q')|\leq |\Pi_{S}(s,p')|$. We can guarantee a ratio not larger than that of $\Pi_{\mbox{orig}}$ as long as the rays $R'$ lie below the turning point $q_{(2i-3,2i-2)}$ of $\Pi_{\mbox{orig}}$. When the ray moves through $q_{(2i-3,2i-2)}$ and $p_{(2i-3,2i-2)}$ 
the bundle of rays (wedge $W_1$ in Figure~\ref{fig-terrain-strat-adapt-analysis}) ends. 
For all rays $R'$ in the wedge $W_1$ we can apply the same argument by comparing $|\Pi_{S}(s,p')|\leq|\Pi_{\mbox{orig}}(s,q')|$. The argumentation holds until $R'$ runs through  $q_{(2i-3,2i-2)}$ and $p_{(2i-3,2i-2)}$  which need not be the same.

Therefore there is another, second wedge $W_2$ of interest for comparisons which actually belong to the boundary worst case, the wedge starts locally at the ray running through $q_{(2i-3,2i-2)}$ and $p_{(2i-3,2i-2)}$  and ends with a vertical ray that passes through~$p_{(2i-3,2i-2)}$, see also  Figure~\ref{fig-terrain-strat-adapt-analysis}. For a ray $R''$ in this wedge and the detection point $p''$ and $q''$  the path length arguments $|\Pi_{S}(s,p'')|\leq|\Pi_{\mbox{orig}}(s,q')|$   still holds but the direct shortest path to $R''$ does not fit to the original strategy any more.  We need another argument for the wedge $W_2$. 

\subsubsection{Decreasing ratios for the rays in $W_2$:} 

\begin{figure}[h]
 \begin{center}
\includegraphics[scale=0.45,page=2]{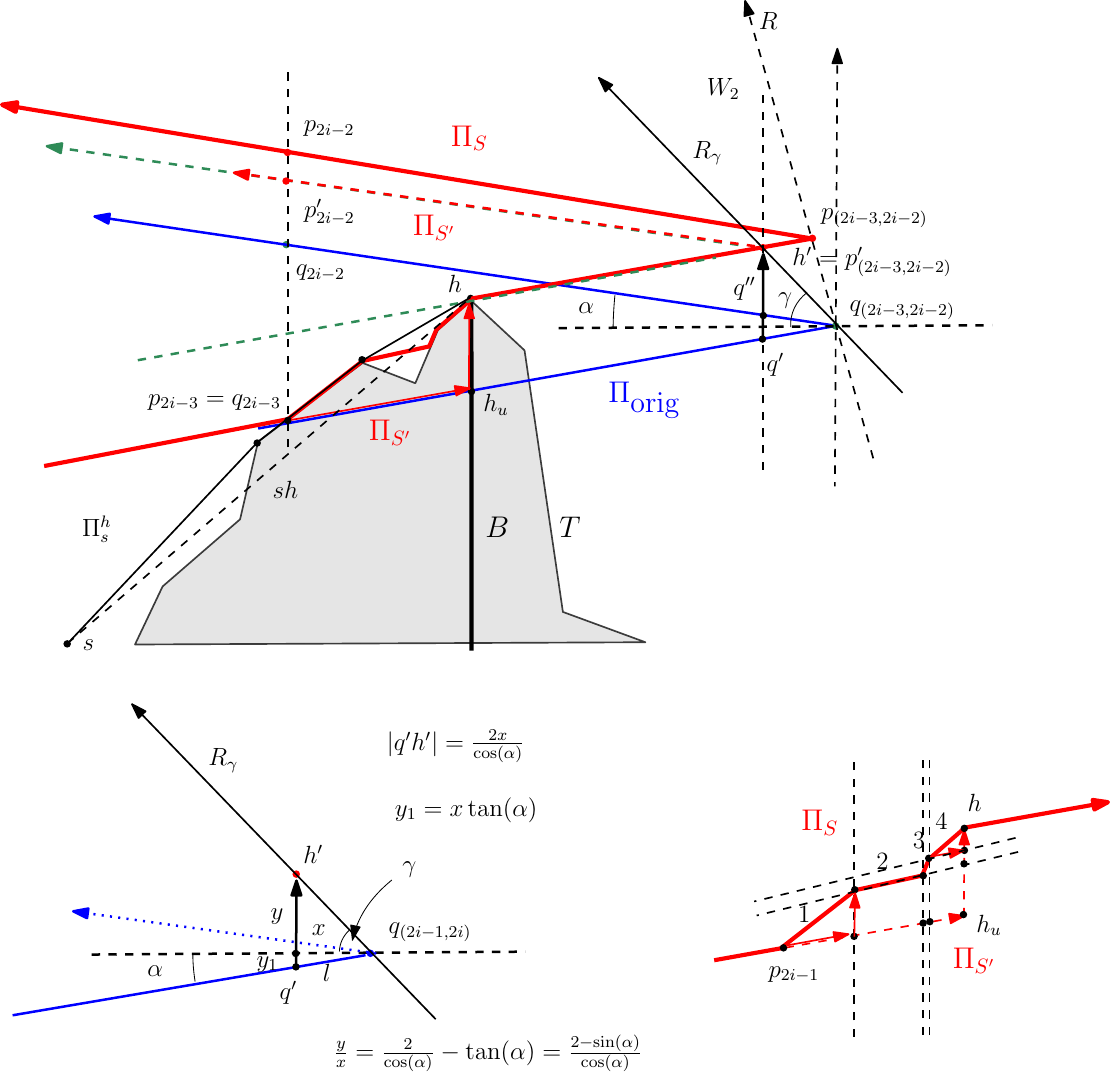}\caption{If the strategy starts moving along the terrain the smallest slope $\gamma$ of the starting ray $R_\gamma$ of $W_2$ will be attained when the terrain just enforces a single upward movement by a vertical barrier $B$. By (successive application of) triangle inequality the length of a straight line movement from $p_{2i-3}$ to $h_u$ and then up to $h$ is always larger than a monotone path of $\Pi_S(p_{2i-3},h)$ enforced by the terrain. By replacing a movement along the terrain correspondingly the return point  $p_{(2i-3,2i-2)}$  moves back to $p'_{(2i-3,2i-2)}=h'$ for $\Pi_{S'}$ and $|h_u h|$ exactly equals $|\Pi_{\mbox{orig}}(q',q'')|$ by design of the strategy. Simple trigonometry shows that the smallest possible slope for $R$ is fixed by $\frac{2-\sin(\alpha)}{\cos(\alpha)}$ and is independent from the length of a segment $|h_u h|$. Thus we have a unique starting ray $R_\gamma$ for the wedge $W_2$.}\label{fig-terrain-strat-adapt-slope}
\end{center}
\end{figure}

The first question is how large can the wedge $W_2$ locally be? If the last turning points $q_{(2i-3,2i-2)}$ and $p_{(2i-3,2i-2)}$  have the the same $X$-coordinate (this happens when we  did not hit the terrain during the last movement into the corresponding direction) the wedge is just the last vertical ray passing through $q_{(2i-3,2i-2)}$ and $p_{(2i-3,2i-2)}$ itself. This gives the same worst case ratio (when the strategy comes back to this coordinate after the return) compared to shortest $X$-distance to this vertical ray, which is the $X$-coordinate of the turning points. 

So how does the wedge $W_2$ change, when the strategy hits the terrain and moves up and what is the effect for the worst case after the next return (the boundary worst case)? We first calculate the minimal possible slope of a starting ray running to some turning points $q_{(2i-3,2i-2)}$ and $p_{(2i-3,2i-2)}$. These two points belong together and we know that 
$|\Pi_{S}(s,p_{2i-2})|=|\Pi_{\mbox{orig}}(s,q_{2i-2})|$ holds for the path length by the design of the strategy. But we have 
$|\Pi_{S}(s,p_{(2i-3,2i-2)})|>|\Pi_{\mbox{orig}}(s,q_{(2i-3,2i-2)})|$  in between, because the strategy climbs upward in comparison to the original strategy! 

So the slope of a ray $R$ passing through $q_{(2i-3,2i-2)}$ and $p_{(2i-3,2i-2)}$ depends on the movements  of 
 $\Pi_{S}(p_{2i-3},p_{(2i-3,2i-2)})$
  against $\Pi_{\mbox{orig}}(q_{2i-3},q_{(2i-3,2i-2)})$.
   Since $q_{(2i-3,2i-2)}$ is always fixed  we consider a first hit point $p_{2i-3}$ and 
 a leave point $h$ of $\Pi_{S}(p_{2i-3},p_{(2i-3,2i-2)})$ which changes 
 $p_{(2i-3,2i-2)}$, see Figure~\ref{fig-terrain-strat-adapt-slope}. By construction we have $|\Pi_{S}(s,p_{2i-3})|=|\Pi_{\mbox{orig}}(s,q_{2i-3})|$ and  $p_{2i-3}$ always have  the same $X$-coordinate as 
 $q_{2i-3}$ but $p_{2i-3}$ never lies below $q_{2i-3}$. Therefore we maximize our chances to 
 minimize the slope of $R$ when $q_{2i-3}$ and  $p_{2i-3}$ are exactly the same points. So let us assume 
 $q_{2i-3}=p_{2i-3}$.
 The slope of~$R$ is given by the $Y$-distance  divided by the $X$-distance of $q_{(2i-3,2i-2)}$ and $p_{(2i-3,2i-2)}$. 
 
  If the movement  $\Pi_{S}(p_{2i-3},h)$ takes away a maximal portion from the budget the point $p_{(2i-3,2i-2)}$ will enforce a minimal slope for~$R$ since the relative position of $p_{(2i-3,2i-2)}$ will change along the $\alpha$-track \emph{back}, see Figure~\ref{fig-terrain-strat-adapt-slope}. By triangle inequality the maximal budget that can be enforced by a movement from $p_{2i-3}$ to the $X$-coordinate of $h$ in $\alpha$-direction to a point $h_u$ on $\Pi_{\mbox{orig}}$ and then a vertical movement moving up to $h$, see $\Pi_S'$ in  Figure~\ref{fig-terrain-strat-adapt-slope}. Thus the worst case minimal slope for  $R$ will also be attained when the maximal budget is used  with a vertical barrier.  And also $h$ can be exactly equal to the return point $p_{(2i-3,2i-2)}$. We indicate this by $h'=p'_{(2i-3,2i-2)}$ in Figure~\ref{fig-terrain-strat-adapt-slope}. We can now precisely calculate the minimal possible starting slope for the wedge $W_2$ for our strategy, independently from the maximal height point $h$. 
 
We consider the vertical line passing through $p'_{(2i-3,2i-2)}=h'$, the line will cross $\Pi_{\mbox{orig}}(q_{2i-3},q_{2i-2})$ at two point $q'$ and $q''$ such that $\Pi_{\mbox{orig}}(q_{2i-3},q'')$ equals  $\Pi_{S}(p_{2i-3},p'_{(2i-3,2i-2)})$ by construction of the strategy w.r.t. using the budget. By the argumentation above the segment $q'h'$ has exact length $|\Pi_{\mbox{orig}}(q',q'')|$ and for the minimal slope we consider the situation given in Figure~\ref{fig-terrain-strat-adapt-slope}.  We have $|q'h'|=2y_1=\frac{2x}{\cos(\alpha)}$ and therefore $y=\frac{2x}{\cos(\alpha)}-y_2=\frac{2x}{\cos(\alpha)}-x\tan(\alpha)$ and the minimal slope is given by $\frac{y}{x}=\frac{2-\sin(\alpha)}{\cos(\alpha)}\approx 1.859957576113$. Alternatively, the angle of the starting ray, say $R_\gamma$, has size $\gamma:=\arctan\left(\frac{2}{\cos(\alpha)}-\tan(\alpha) \right)\approx 1.0774868815$  (which is $\gamma\approx 61.73545079^{\circ}$). At this stage the precise values are only used to give an impression of the slopes.

Overall, locally for a corresponding  rotational point $p_{(2i-3,2i-2)}$ and the corresponding  $q_{(2i-3,2i-2)}$ the wedge $W_1$ runs at least to angle $\gamma$ and the wedge $W_2$ does not start earlier than $\gamma$ and goes to $\pi/2$. 
Any return point $p_{\gamma}$ on the ray $R_{\gamma}$ can serve as a worst case starting rotational  point for the wedge $W_2$ if the budget was maximally used by a vertical barrier. Any other movement on the terrain will result in a return point above $R_{\gamma}$. 

Finally, we would like to argue that the boundary maximum of the original strategy for $\beta=\pi/2$ subsumes any local worst case for rays in a wegde $W_2$ also. We will show that the overall worst case for $W_2$ is given by the last vertical ray that rotates around $p_{(2i-3,2i-2)}$.  

As already shown above, the worst case for the straetgy and the wedge $W_2$ will be attained when the return point $p_{(2i-3,2i-2)}$ lies on the ray $R_{\gamma}$ (for $\gamma\approx 61.73545079^{\circ}$ calculated above) that runs through the return point $q_{(2i-3,2i-2)}$ of the original strategy. Thus we can actually move this point $p_{(2i-3,2i-2)}$ up along the ray $R_{\gamma}$ starting from $q_{(2i-3,2i-2)}$. 

Additionally, the location of the current maximal height point $h$ on the movement to $p_{(2i-3,2i-2)}$ will influence the shortest path from $s$ to the rays in the wedge $W_2$. Assuming that a vertical barrier running through $h_u h$ will be used also helps to make this path as short as possible. Therefore for the overall worst case for the ratio technically we only have to consider a vertical segment  $h_u h$ on the original strategy whose length fixes the return point $p_{(2i-3,2i-2)}$ on $\Pi_S$ and also the shortest path to any ray $R$ in $W_2$ that rotates around $p_{(2i-3,2i-2)}$ for angles from $\gamma$ to $\pi/2$.

Interestingly, if we move the segment $h_u h$ close enough to the return point  $p_{(2i-3,2i-2)}$  and $h_u h$ is also small enough, the shortest path to the starting ray $R_{\gamma}$ can be a direct segment from $s$ that is perpendicular to $R_{\gamma}$. If we rotate the rays further on the shortest possible path will finally run over $p_{(2i-3,2i-2)}$. To show that this behaviour is given we can have a look at a line passing through all original return points $q_{(2i-3,2i-2)}$. 

By design of the strategy in Section~\ref{subsect-design} the turning points of $\Pi_{\mbox{orig}}$ were previously precisely calculated by 
$$q_{i-1}=\left((-1)^{i-1}\cdot r^{i-1},2\tan(\alpha)\left(\frac{r^{i}-1}{r-1}-\frac{1}{2}r^{i-1}\right)\right)$$ starting with $i=1$ and then $i=3,5,7,\ldots$
for the turns on the right side and $i=2,4,6,8,\ldots$ for the left side turns. 
In our new notation we can directly translate the notion into 
$$q_{(2i-3,2i-2)}=\left((-1)^{i-1}\cdot r^{i-1},2\tan(\alpha)\left(\frac{r^{i}-1}{r-1}-\frac{1}{2}r^{i-1}\right)\right)$$ with the same interpretation for left and right.

We consider the ray $R_{\gamma}$ that runs through some $q_{(2i-3,2i-2)}$ and calculate the shortest distance to the ray from the start~$s$.
Note that we have already considered and calculated such distances against general rays with angles~$\beta$ in  Formula~(\ref{equ-shortestpath}) of Section~\ref{subsect-design}. Thus, the distance to $R_{\gamma}$ is given by an orthogonal line segment of length 

$$|s\,t_\gamma|=\cos(\gamma)\cdot \left(\tan(\gamma)\,r^{i-1}+2\tan(\alpha)\left(\frac{r^{i}-1}{r-1}-\frac{1}{2}r^{i-1}\right)\right)$$

where $\tan(\gamma)=\frac{2}{\cos(\alpha)}-\tan(\alpha)$ holds which results in 
\begin{eqnarray}
\cos(\gamma)\cdot \left(\frac{2}{\cos(\alpha)}\,r^{i-1}+2\tan(\alpha)\left(\frac{r^{i}-1}{r-1}-r^{i-1}\right)\right)& = &\nonumber \\
2\cos(\gamma)\cdot\left( \frac{1-\sin(\alpha)}{\cos(\alpha)}+\tan(\alpha)\frac{r-\frac{1}{r^{i-1}}}{r-1}\right)r^{i-1}\label{eqn-dist-R-gamma}
\end{eqnarray}



First, we compare this length to the shortest distance of the boundary worst case ray $R_{\frac{\pi}{2}}$ of $\Pi_{\mbox{orig}}$ which is $r^{i-1}$. Taking all our precise values into account the factor in front of $r^{i-1}$ in (\ref{eqn-dist-R-gamma}) is significantly larger than $1$  for any $i\geq 2$. For the starting ray $R_\gamma$ the boundary worst case of the original strategy for $\beta=\pi/2$ is always worse! Note that here we have used our strategy values of $r$ and $\alpha$ for an analysis for the first time! 

Starting from $R_{\gamma}$ in $W_2$ the rays now rotate around $p_{(2i-3,2i-2)}$ of the strategy $\Pi_S$
rather than originally around $q_{(2i-3,2i-2)}$ and the distance might shrink in comparison to $r^{i-1}$. The precise location of the current worst case $p_{(2i-3,2i-2)}$ on $R_\gamma$ depends on the point $h$ of the terrain that has maximal height on the movement toward  $p_{(2i-3,2i-2)}$. 
As mentioned above it is worse for the ratios of the strategy when the terrain is just a single barrier and the movement to $h$ is a vertical segment. For any given $h$, the return point is precisely given. But we have some room for the location of $h$ which influences the shortest path to the rays in $W_2$. 

\begin{figure}[h]
 \begin{center}
\includegraphics[scale=0.45,page=2]{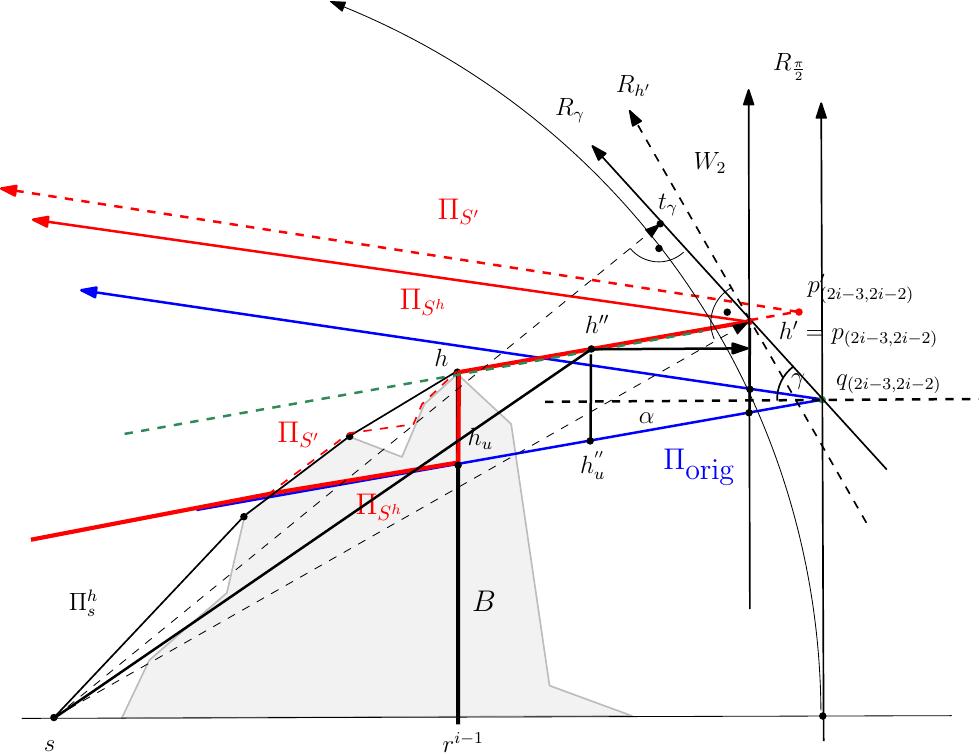}\caption{The ratio calculations for the wedge $W_2$. The strategy meets a highest point $h$ locally, see $\Pi_{S'}$. As a local worst case we can first assume that a single vertical barrier at $h$ takes more budget away. This   leads to strategy $\Pi_{S^h}$ which also translates the return and rotational point $p_{(2i-3,2i-2)}$ closer to the start and also leads to shorter distances from the start over the terrain to the rays in the corresponding wedge $W_2$. For the starting ray $R_\gamma$ we know that the original boundary worst case has a larger ratio, the shortest path segment $st_\gamma$ from $s$ to  $R_\gamma$ is always larger than the shortest path length $r^{i-1}$ from $s$  to the vertical ray $R_{\frac{\pi}{2}}$ through $q_{(2i-3,2i-2)}$. This argument remains valid for the rays in $W_2$ at least up to a ray $R_{h'}$ whose orthogonal shortest path from the start is just the segment $s\,p_{(2i-3,2i-2)}$. We know that  $|s\,p_{(2i-3,2i-2)}|>s\,t_\gamma$ and just apply the same argument. After that any shortest path to a ray in $W_2$ has to visit the height point first. The remaining case is that we are looking for a worst case passing a height point $h''$ and the worst case is also given by a vertical ray passing through a rotational point $p_{(2i-3,2i-2)}$~on $R_\gamma$.}\label{fig-terrain-strat-adapt-slope-W2}
\end{center}
\end{figure}

We have a situation as depicted in Figure~\ref{fig-terrain-strat-adapt-slope-W2}. The ray $R_y$ is the same for any $h$ but the shortest path from the start to the fixed ray $R_y$ (i.e., a segment $s\,t_\gamma$ orthogonal to $R_\gamma$)  does not necessarily cross the barrier with top most point $h$.  This depends on the location of $h$ and the length of  $h_u h$. We can \emph{translate} $h$ and the barrier forward up to the return point, say we use $h'=p_{(2i-3,2i-2)}$. The segment $s\,h'=s p_{(2i-3,2i-2)}$ now is indeed barrier free. The shortest segment $s\, t_\gamma$ from $s$   to $R_\gamma$  can hit $R_\gamma$ above $h'$ and can also be obstacle free. This just depends on the length $h_u h$ of the current barrier segment.  We already argued that $s\,t_\gamma$ is always larger than the length $r^{i-1}$ for the boundary worst case of the original strategy. The adapted strategy will also find $R_\gamma$ on the next way back earlier. The original strategy has larger ratios. 

The rays of interest starts rotating around $p_{(2i-3,2i-2)}$. 
The above argumentation remains at least valid  until 
the segment $s\,p_{(2i-3,2i-2)}$ gets orthogonal to a first ray $R_{h'}$ in $W_2$. Note that $p_{(2i-3,2i-2)}$ was on $R_\gamma$ and  $s\,p_{(2i-3,2i-2)}$ is even larger than  $s\,t_\gamma$ and is again subsumed by the original boundary worst case. If we rotate the rays further on around $p_{(2i-3,2i-2)}$ the shortest path to the remaining rays in $W_2$ now definitely have to pass a given barrier height point. We can still make use of a worst case translated height point $h''$ to be on the very safe side. Nevertheless for the remaining rays of this kind the overall worst case  in $W_2$ will be the final vertical boundary ray  passing through $p_{(2i-3,2i-2)}$. Obviously the path length to this final ray for a path that has to meet some height point $h''$ first is minimized for this boundary ray. 
And this vertical boundary ray will also finally be detected last in the backward movement, i.e., the maximal path length for detecting all rays in $W_2$ will be achieved for a final vertical ray passing $p_{(2i-3,2i-2)}$.   

Now it is sufficient to claim that for any such  
translation point $h''$, the shortest path from $s$ to $h''$ and then (horizontally) toward the current final vertical worst case ray that passes through $p_{(2i-3,2i-2)}$ is always larger than $r_{i-1}$. If this holds, by the argumentation above the ratios for the rays in $W_2$ are all subsumed by the original boundary worst case as shown before. 

\begin{figure}[h]
 \begin{center}
\includegraphics[scale=0.45,page=2]{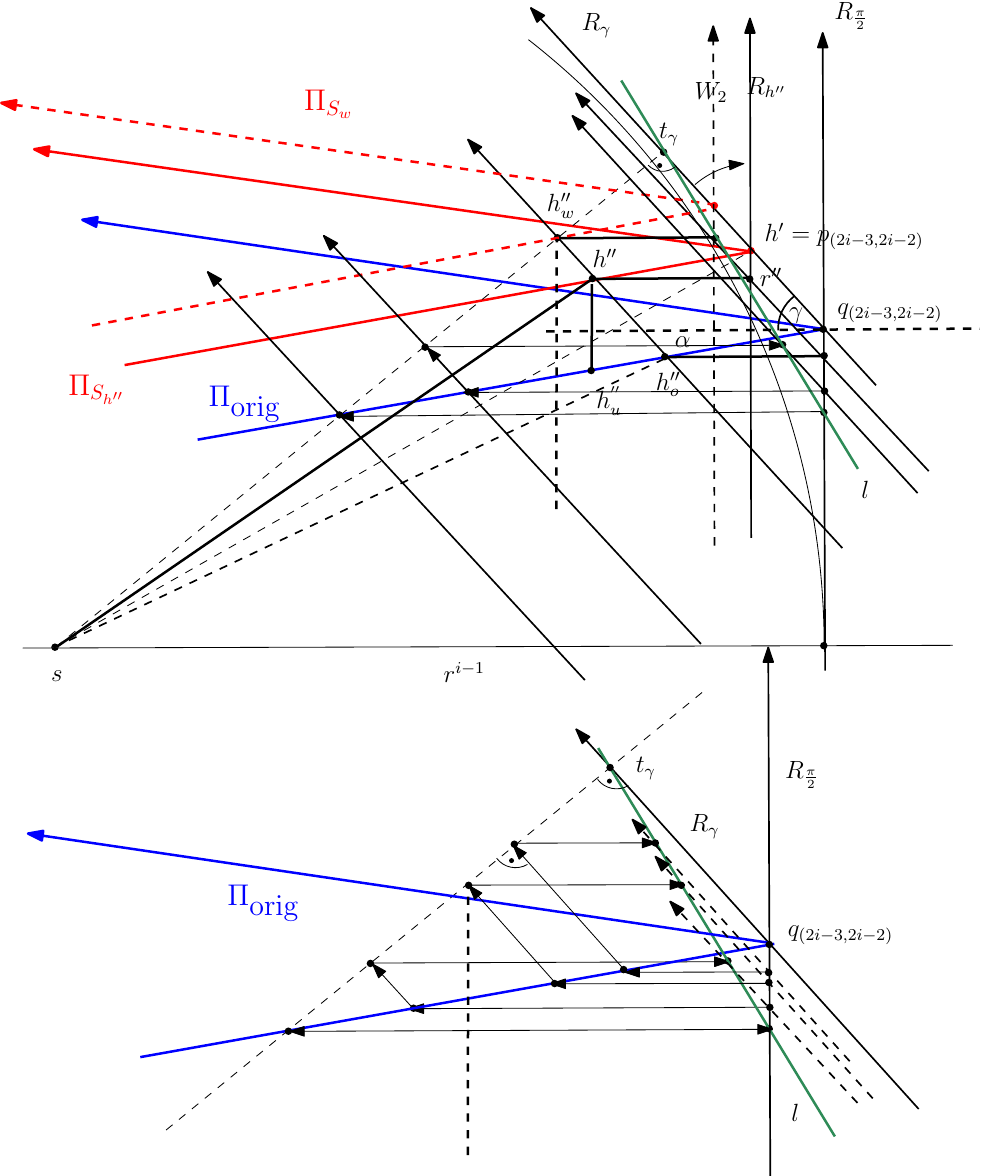}\caption{We can find a worst case description for the final rays in $W_2$. The worst case height points $h''_w$ for the strategy are finally located on $s\,t_\gamma$, i.e., the shortest path from $s$ to $R_\gamma$. In this case the movement to the height point on $s\,t_\gamma$ and then horizontally to the vertical ray of the corresponding return point is minimized for a given fixed distance of  $h''\,r''$.  Fortunately, all such segment translations have the final end point on a common line. This line $l$ runs through the end point of $s\,t_\gamma$ (for horizontal distance $0$) and intersects the original worst case ray $R_{\frac{\pi}{2}}$ at the $Y$-coordinate of the intersection of the original strategy with $s \, t_\gamma$. In this case the final horizontal distance in the shortest path is maximal and just given by the horizontal segment between the intersection points.  
 }\label{fig-terrain-strat-adapt-slope-W2-proof}
\end{center}
\end{figure}

We consider the given situation isolated in Figure~\ref{fig-terrain-strat-adapt-slope-W2-proof}. 
Let us fix the vertical distance from such a height point $h''$ to the corresponding vertical ray $R_{h''}$ that runs through the return point $h'=p_{(2i-3,2i-2)}$. 
The vertical orthogonal segment from $h''$  to ray $R_{h''}$ hits the ray at point $r''$. What is the overall worst case for this given distance $|h''\,r''|$ w.r.t. the shortest path to such a ray? We can translate the segment $h''\,r''$ along a line that is parallel to $R_\gamma$ and runs through~$r''$. If we translate this segment down to the original strategy~$\Pi_{\mbox{orig}}$, we can already be sure that the movement from $s$ to the image of $h''$ combined with the image of the segment $h''\,r''$ is larger than  $r_{i-1}$. 
This is also a reasonable case when the barrier ends exactly at the original strategy. Obviously the overall worst case for a translated image of segment $h''\,r''$ is given if we translate the segment such that the image of $h''$ meets the shortest path segment $s\,t_\gamma$ from $s$ to $R_\gamma$ at some point $h''_w$. In this special case the shortest path from $s$ to the image  of $h''$ passing along the parallel line of $R_y$ passing  $h''$  in addition with the fixed path length of $h''\,r''$ is minimized. It suffices to show that all such translated paths for any such given length  $h''\,r''$ are shorter than $r_{i-1}$. 

\begin{figure}[h]
 \begin{center}
\includegraphics[scale=0.5]{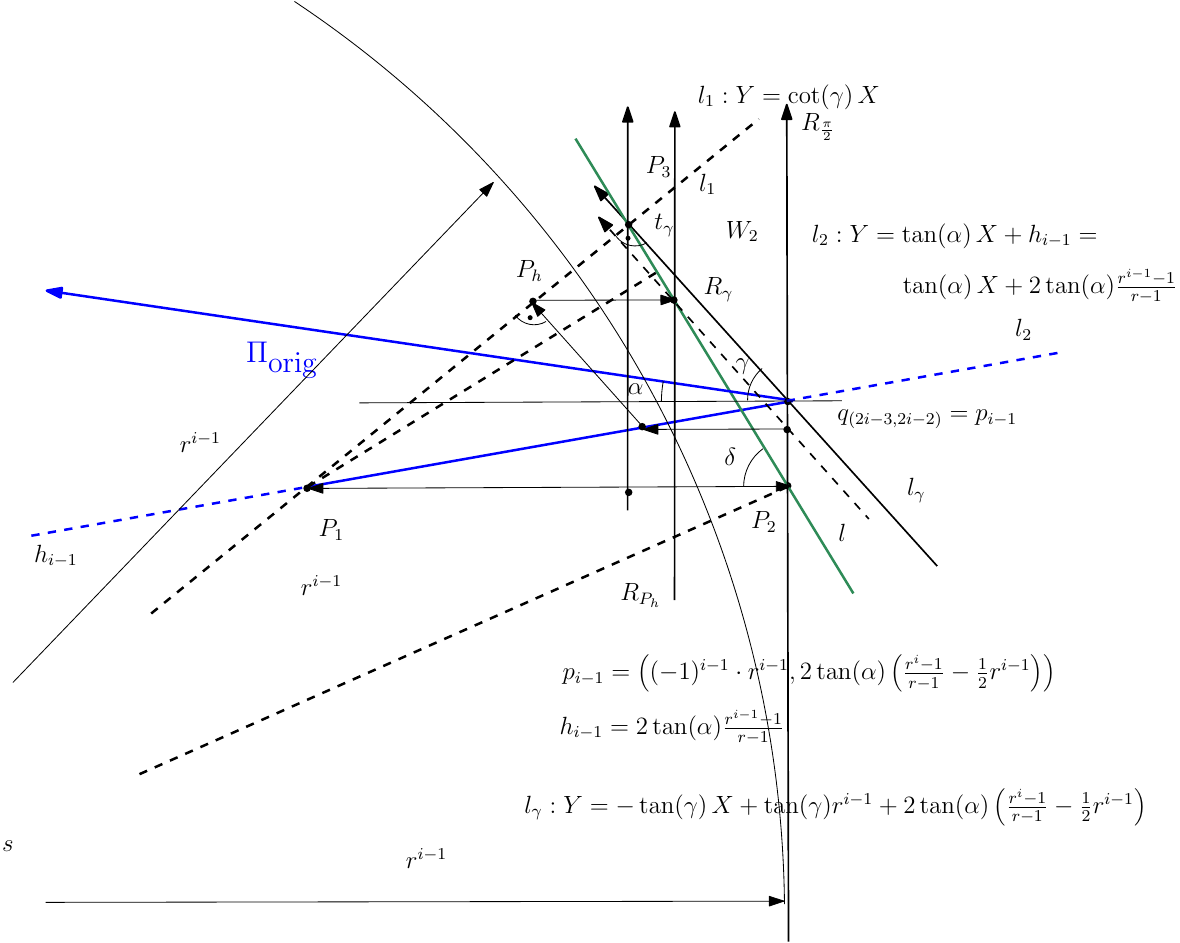}\caption{For the precise calculation indicated in Figure~\ref{fig-terrain-strat-adapt-slope-W2-proof} we set up the lines $l_2$ (as part of the strategy) and $l_1$ (the line passing through $s\, t_\gamma$). We can compute the intersection points $P_1$ of $l_1$ and $l_2$ and $P_3$ of $l_\gamma$ and $l_1$ and also find the $Y$-coordinate of $P_3$. So we can calculate the slope angle $\delta$ of $l$ by $P_2$ and $P_3$.  If the length of a path from $s$ to some $P_h$ plus the length of the horizontal path from $P_h$ to $l$ (this is the minimal path length to a vertical ray $R_{P_h}$) is always targer than $r^{i-1}$ the worst case ratio for the original strategy for the ray $R_{\frac{\pi}{2}}$ (visited by the backward movement of $\Pi_{\mbox{orig}}$) is larger than any vertical worst case ray $R_{P_h}$ that stems from the wedge $W_2$.}\label{fig-final-triangle-proof}
\end{center}
\end{figure}

Note that for any fixed $h''\,r''$  the translation of the segment and also of its end point is a linear transformation. 
As indicated in the lower part of Figure~\ref{fig-terrain-strat-adapt-slope-W2-proof} this means that for any possible distance,  the final image points of $r''$ all lie on the same line~$l$. Also note that it is indeed possible to trigger any such path to a corresponding ray, by making use of a barrier that precisely ends at a point on $s\,t_\gamma$ but above the original strategy! So finally for the evaluation of the rays in $W_2$ we only have to consider paths represented in a triangle that depends on the point $q_{(2i-3,2i-2)}$ of the original strategy with our fixed parameters of the strategy! Therefore finally we exactly consider any such triangle as given in Figure~\ref{fig-final-triangle-proof}. It represents all (final) movements of all worst case situation.  

Let 
 
 $$
l_1: Y=\cot(\gamma)\,X=\frac{\cos(\alpha)}{2-\sin(\alpha)}\, X$$
 be the line passing through the segment $s\,t_\gamma$ and $$l_2: Y=\tan(\alpha)\,X + h_{i-1}=
 \tan(\alpha)\,X + 2\tan(\alpha)\frac{r^{i-1}-1}{r-1}$$ 
 be the line passing to the forward movement of the original strategy.  We locally (dependent on $i$) denote the line passing $R_\gamma$ by  $l_\gamma$. For the strategy design we had already computed all such rays (for any $i$) in Section~\ref{subsect-design}, therefore we have 

$$l_\gamma: Y=-\tan(\gamma)\,X+\tan(\gamma)\,r^{i-1}+2\tan(\alpha)\left(\frac{r^{i}-1}{r-1}-\frac{1}{2}r^{i-1}\right)\,.$$

We would like to compute the endpoints $P_1$, $P_2$  and  $P_3$ of the triangle in order to find the line $l$ passing $P_2$ and $P_3$.  
Here $t_\gamma=P_3$ holds and  $P_1$ is the  intersection of $l_1$ and $l_2$. The point $P_3$ is the intersection point of the vertical (local) worst case ray $R_{\frac{\pi}{2}}$ and the horizontal line passing $P_1$. 

By simple calculations the intersection $P_1=(P_1(X),P_1(Y))$ of $l_1$ and $l_2$ gives

$$
P_1(X) =  \left(\frac{2\tan(\alpha)}{\cot(\gamma)-\tan(\alpha)}\right)\frac{r^{i-1}-1}{r-1}
		= \left(\frac{2\sin(\gamma)\sin(\alpha)}{\cos(\gamma+\alpha)}\right)\frac{r^{i-1}-1}{r-1}\nonumber
$$ 

and 

$$ P_1(Y)  =  \left(\frac{2\cot(\gamma)\tan(\alpha)}{\cot(\gamma)-\tan(\alpha)}\right)\frac{r^{i-1}-1}{r-1}
 =  \left(\frac{2\cos(\gamma)\sin(\alpha)}{\cos(\gamma+\alpha)}\right)\frac{r^{i-1}-1}{r-1}$$



 and we also have $$P_2=\left(r^{i-1},\left(\frac{2\cos(\gamma)\sin(\alpha)}{\cos(\gamma+\alpha)}\right)\frac{r^{i-1}-1}{r-1}\right)\,.$$

Furthermore the intersection $P_3=(P_3(X),P_3(Y))$ of $l_1$ and $l_\gamma$ gives 
\begin{eqnarray*}
P_3(X)& = & \frac{\tan(\gamma)}{\cot(\gamma)+\tan(\gamma)}r^{i-1}+\frac{2\tan(\alpha)}{\cot(\gamma)+\tan(\gamma)}\left(\frac{r^{i}-1}{r-1}-\frac{1}{2}r^{i-1}\right)\\
& = &
 \sin(\gamma)^2\, r^{i-1}+2\sin(\gamma)\cos(\gamma)\tan(\alpha)\left(\frac{r^{i-1}-1}{r-1}+\frac{1}{2}r^{i-1}\right)
\end{eqnarray*}

and

$$P_3(Y)=\sin(\gamma)\cos(\gamma)\,r^{i-1}+2\cos(\gamma)^2 \tan(\alpha)\left(\frac{r^{i-1}-1}{r-1}+\frac{1}{2}r^{i-1}\right)\,.$$ 

Computing the slope of $l$ is given by $\frac{P_2(Y)-P_3(Y)}{P_2(X)-P_3(X)}$ which gives



$$\frac{2\cos(\gamma)\left(\frac{\sin(\alpha)}{\cos(\gamma+\alpha)}-\cos(\gamma)\tan(\alpha)\right)\frac{r^{i-1}-1}{r-1}-\cos(\gamma)\left(\sin(\gamma)+\cos(\gamma)\tan(\alpha)\right)r^{i-1}}{-2\sin(\gamma)\cos(\gamma) \tan(\alpha)\frac{r^{i-1}-1}{r-1}+\left(1-\sin(\gamma)^2-\sin(\gamma)\cos(\gamma) \tan(\alpha)\right)r^{i-1}}\,.$$

We can now divide the enumerator and the denominator by $r^{i-1}$, respectively. For convenience, we can get rid of the factors $\frac{1}{r^{i-1}}$ in the enumerator and  in the denominator, both will decrease in $i$ and rapidly went to zero. Then we multiply enumerator and denominator by $(r-1)$ and  the (asymptotic) slope of any $l$ is thus given by 



$$\frac{2\cos(\gamma)\left(\frac{\sin(\alpha)}{\cos(\gamma+\alpha)}-\cos(\gamma)\tan(\alpha)\right)\frac{1}{r-1}-\cos(\gamma)\left(\sin(\gamma)+\cos(\gamma)\tan(\alpha)\right)}{-2\sin(\gamma)\cos(\gamma) \tan(\alpha)\frac{1}{r-1}+\left(1-\sin(\gamma)^2-\sin(\gamma)\cos(\gamma) \tan(\alpha)\right)}$$ 

and multiplication by $(r-1)$ on both levels simplifies to 



$$\frac{
2\cos(\gamma)\frac{\sin(\alpha)}{\cos(\gamma+\alpha)}
-\cos(\gamma)^2\tan(\alpha)+\cos(\gamma)\sin(\gamma)
-\cos(\gamma)\left(\sin(\gamma)+\cos(\gamma)\tan(\alpha)\right)r}
{-\sin(\gamma)\cos(\gamma) \tan(\alpha)-1+\sin(\gamma)^2+\left(1-\sin(\gamma)^2-\sin(\gamma)\cos(\gamma) \tan(\alpha)\right)r}$$

which further simplifies to 


$$\frac{
\cos(\gamma)^2\tan(\alpha)(r+1)+\cos(\gamma)\sin(\gamma)(r-1)-2\cos(\gamma)\frac{\sin(\alpha)}{\cos(\gamma+\alpha)}
}
{\sin(\gamma)\cos(\gamma)\tan(\alpha)(r+1)+(\sin(\gamma)^2-1)(r-1)}\,.$$ 

Making use of the values we have calculated before the above slope, say $s_l$,  can be calculated as 
 $s_l \approx -2.9506286930748570895843840844464$ or alternatively the angle $\delta$ at $P_2$ expanded between the segment $P_1\,P_2$ and $l$ (see Figure~\ref{fig-final-triangle-proof}) can be calculated as $\delta\approx 71.277924040892699508731311652809^\circ$. Note that we did all calculations by Matlab and with very high precision and the numerical numbers just give us an impression. 

\begin{figure}[h]
 \begin{center}
\includegraphics[scale=0.45]{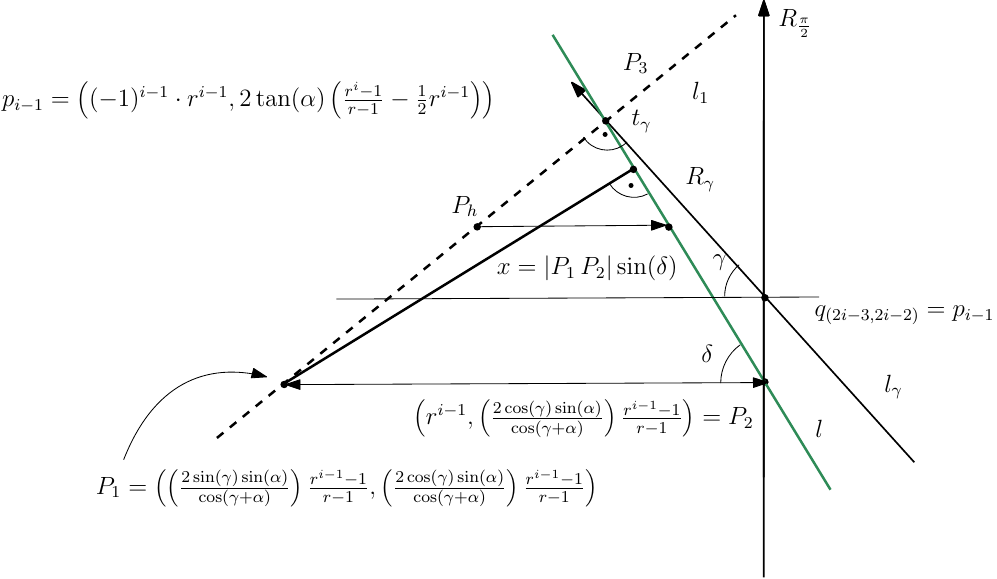}\caption{ For the final outcome we compare the length of any path from the start $s$ to a point $P_h$ and then vertically prolonged to meet $l$. This path should be longer than  $r^{i-1}$.  The shortest path from $P_1$ to $l$ calculated by the law of sine has length $x=\sin(\delta)|P_1\,P_2|$. If we consider the path from $s$ to $P_1$ and add the distance $x$ we obtain a lower bound on all path that have to be considered. Fortunately, with our calculated values we can show that this lower bound is always larger than $r^{i-1}$ for any~$i\geq 2$. Note that the figure represents a realistic scenario since $\delta\approx 71.277924 $ and $\gamma\approx 61.73545$ have been calculated! }\label{fig-final-simple-triangle-proof}
\end{center}
\end{figure}

For the final comparison we would like to make use of these almost precise values. We calculate the length of the path from the start to $P_1$ and additionally we consider the shortest path from $P_1$ to $l$ which meets $l$ in a perpendicular way, see Figure~\ref{fig-final-simple-triangle-proof}. 
The length of this overall path is a lower bound of the length of any terrain induced path that run from $s$ to a point $P_h$ and then horizontally toward $l$. 
So if the sum of these two distances is already  larger than the distance of shortest path from $s$ to the ray $R_{\pi/2}$ (which is $r^{i-1}$) we are done. In this case the pure strategy attains a larger ratio for the local worst case ray $R_{\pi/2}$ against any other worst case situation given by the terrain. 

By simple calculations we have 

$$|s\,P_1|=\left(\frac{2\sin(\alpha)}{\cos(\gamma+\alpha)}\right)\frac{r^{i-1}-1}{r-1}$$

and $$|P_1\,P_2|=r^{i-1}-\left(\frac{2\sin(\gamma)\sin(\alpha)}{\cos(\gamma+\alpha)}\right)\frac{r^{i-1}-1}{r-1}\,.$$

Now finally we have to show that 
$\sin(\delta)|P_1\,P_2|+ |s\,P_1|$ is larger than $r^{i-1}$. We calculate 
$\sin(\delta)|P_1\,P_2|+ |s\,P_1| - r^{i-1}$ and have to show 

$$(\sin(\delta)-1) r^{i-1}+(1-\sin(\gamma)\sin(\delta))\left(\frac{2\sin(\alpha)}{\cos(\gamma+\alpha)}\right)
\frac{r^{i-1}-1}{r-1}$$

is larger than zero. Thus we consider 

\begin{eqnarray*}
(1-\sin(\delta)) r^{i-1} & < & (1-\sin(\gamma)\sin(\delta))\left(\frac{2\sin(\alpha)}{\cos(\gamma+\alpha)}\right)
\frac{r^{i-1}-1}{r-1}\\
(1-\sin(\delta)) & < & (1-\sin(\gamma)\sin(\delta))\left(\frac{2\sin(\alpha)}{\cos(\gamma+\alpha)}\right)\left(\frac{1}{r-1}-\frac{1}{r^{i-1}}\right)\\
(1-\sin(\delta))(r-1) & < & (1-\sin(\gamma)\sin(\delta))\left(\frac{2\sin(\alpha)}{\cos(\gamma+\alpha)}\right)\left(1-\frac{r-1}{r^{i-1}}\right)\,.
\end{eqnarray*}

For the final calculation we can make use of a (slightly) but significantly smaller $\delta$ which decrease our chances to fulfil the last inequality due to the geometric interpretation in Figure \ref{fig-final-simple-triangle-proof}. Using a significantly smaller $\delta$ also simply subsumes our asymptotic calculation of the slope (leaving out the factors $\frac{1}{r^{i-1}}$) before. 
By just making use of an $\delta'=71$ (significantly smaller than the calculated value) we are on the very safe side because $\sin(\delta)$ and in turn the distance $x$ shrinks. This reduces our chances to fulfil the inequality. Using all other so far calculated values we can still conclude that 
 
$$(1-\sin(\gamma)\sin(\delta'))\left(\frac{2\sin(\alpha)}{\cos(\gamma+\alpha)}\right)\left(1-\frac{r-1}{r^{i-1}}\right)-(1-\sin(\delta'))(r-1)>0.1$$ 
holds already for $i=5$. Actually already for $i=2$ the left hand side value is larger than $0.0339$. For larger $i$ the value will just increase, asymptotically to a value larger than $0.1195172$ for $\delta=71.277$. Here we can make use of the almost precise values for the second time. We are far away from precision problems w.r.t. our claims. 

Altogether, although we work with some approximated values it is shown that w.r.t. our strategy design, the ratio that is attained at the terrain problem will never be worse than the ratio of the pure strategy without the terrain.

\begin{figure}[h]
 \begin{center}
\includegraphics[page=2,scale=0.4]{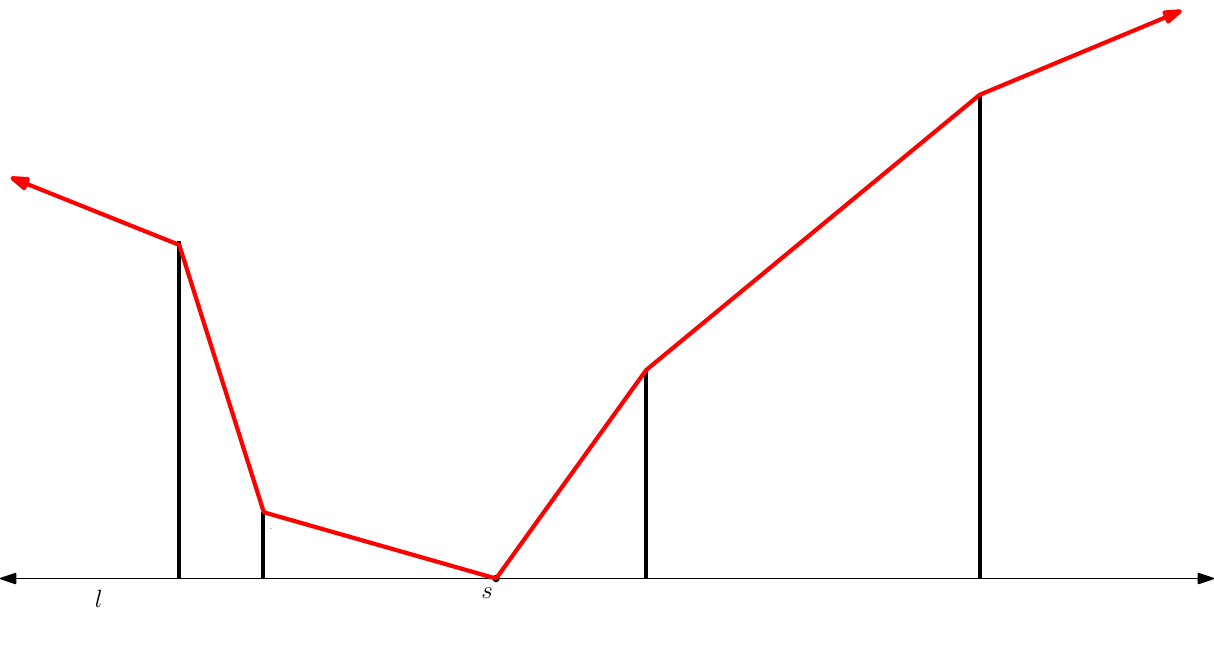}\caption{For a reasonable online strategy design, for a worst case we can assume that the terrain consists of vertical barriers with increasing heights, respectively.}\label{fig-core}
\end{center}
\end{figure}

Finally, we would like to summarize that the terrain generally can be simplified, at least for our kind of cow-path like strategies.   
\subsection{The core of the terrain problem}\label{sect-core}

\subsubsection{worst case is attained by simple vertical  barriers:}

Figure~\ref{fig-terrain-strat-adapt-slope} already  locally exemplifies a movement along the terrain and its impact on the ratios for searching the rays. If we follow the terrain for a while into one direction we will finally meet a highest level point $h$ on the terrain before we return to the other side. Replacing the terrain part by a vertical line segment barrier $B$ with topmost point $h$ can only be worse for the strategy w.r.t. ratios for rays detected in the future. The geometric argumentation is shown in Figure~\ref{fig-terrain-strat-adapt-slope}. Note that $B$ is already part of the original terrain. Simple vertical barriers will decrease the length to the shortest path to a ray and also enforce a maximal use of the budget for moving up along the terrain.

Generally, this means that for our strategy we actually can assume that we only have to cope with vertical line segment barriers on both sides and since there is no need to move down for any reasonable strategy, we can also assume that the height of the barriers increase successively on both sides, respectively.  We can expect an online situation as indicated in Figure~\ref{fig-core}.

\section{Conclusion and future work}\label{sect-concl}

We considered the problem of searching for an unknown ray in the half-plane such that the shortest path to the unknown ray also runs inside this half-plane. A generalized and optimized cow-path guarantees a competitive ratio not larger than $9.12725$ and there is also a lower bound of $9.06357$ for any strategy. To this end for the strategy we balanced out two worst case rays for the strategy, one vertical ray and one ray of slope $\beta$ that runs in tangent to the strategy shortly after a turn. The problem was motivated by a terrain search problem that was recently considered. We show that we can adjust our strategy to the terrain search problem and by geometric arguments and the parameters of our strategy we show that the ratio can only decrease if the terrain is present. The consideration also somehow isolates the core of the terrain search problem by argueing that only vertical barriers of increasing heights have to be considered. This gives rise for obtaining the optimal competitive ratio for the problem. The current gab between upper and lower bound is less than~$0.06368$.

%

 \bibliographystyle{splncs04}
 \bibliography{references,g}

\end{document}